\documentclass{lmcs} 
\pdfoutput=1
\usepackage[utf8]{inputenc}

\usepackage{lastpage}
\lmcsdoi{19}{2}{14}
\lmcsheading{}{\pageref{LastPage}}{}{}%
{Jan.~28,~2022}{Jun.~06,~2023}{}


\usepackage{hyperref}
\usepackage{cleveref}
\usepackage{xspace}
\usepackage{thm-restate}

\newcommand\myitem[1]{(\hyperref[item:#1]{\ref*{item:#1}}{})\xspace}

\def\marker{\bar}
\def\bu{{\marker u}}
\def\bv{{\marker v}}

\def\bx{{\marker x}}
\def\by{{\marker y}}

\def\Nesetril{Ne\v{s}et\v{r}il\xspace}
\def\Dvorak{Dvo\v{r}\'{a}k\xspace}
\def\Kral{Kr\'{a}l\nobreak\hspace{-.12em}'\xspace}
\def\Gajarsky{Gajarsk\'{y}\xspace}

\def\phi{\varphi}
\def\cal{\mathcal}
\def\N{\mathbb N}
\def\epsilon{\varepsilon}

\newcommand\tw{{\rm tw}}
\newcommand\td{{\rm td}}

\newcommand\reach{{\rm Reach}}
\newcommand\wreach{{\rm WReach}}
\newcommand\col{{\rm col}}
\newcommand\wcol{{\rm wcol}}
\def\G{\mathcal G}

\newcommand{\tp}{\textnormal{tp}_q}
\newcommand{\tpp}[1]{\textnormal{tp}_{#1}}
\def\typi{\textit{type}}

\begin{document}

\title[Lacon-, Shrub- and Parity-Decompositions]{Lacon-, Shrub- and Parity-Decompositions: Characterizing Transductions of Bounded\texorpdfstring{\\}{ }Expansion Classes}
\titlecomment{An extended abstract of this paper has appeared in the proceedings of the 36th Annual ACM/IEEE Symposium on Logic in Computer Science (LICS 2021).}

\author[J.~Dreier]{Jan Dreier\lmcsorcid{https://orcid.org/0000-0002-2662-5303}}

\address{Algorithms and Complexity Group, TU Wien, Austria}	
\email{dreier@ac.tuwien.ac.at}  






\begin{abstract}
  \noindent The concept of bounded expansion provides a robust way to capture sparse graph classes with interesting algorithmic properties.
Most notably, every problem definable in first-order logic can be solved in linear time on bounded expansion graph classes.
First-order interpretations and transductions of sparse graph classes lead to more general, dense graph classes that 
seem to inherit many of the nice algorithmic properties of their sparse counterparts.

In this paper, we show that one can encode graphs from a class with structurally bounded expansion
via lacon-, shrub- and parity-decompositions from a class with bounded expansion.
These decompositions are useful for lifting properties from sparse to structurally sparse graph classes.

\end{abstract}

\maketitle

\section{Introduction}

Many hard graph problems become easier if we assume the input to be \emph{well-structured}.
Usually, this is enforced by assuming that the input belongs to a certain kind of (infinite) graph class.
This can, for example, be the class of all planar graphs or the class of all graphs with maximal degree at most $d$ for some number $d$.
Many of the established tractable graph classes are \emph{sparse} in the sense that their graphs have relatively few edges compared to their number of possible edges.

\emph{Algorithmic meta-theorems} state that whole families of problems become more trac\-table
if we restrict the input to certain graph classes.
The best known example is Courcelle's theorem~\cite{Courcelle1990}, which states that every problem definable in monadic second-order logic
can be solved in linear time on bounded treewidth classes.
It has been shown in a series of papers that all problems definable in first-order logic can be solved in linear time on 
bounded degree~\cite{Seese1996},
excluded minor~\cite{FlumG2001}, 
locally bounded treewidth~\cite{FrickG2001}
and more general sparse graph classes~\cite{dvorak2010deciding,GroheKS2017}.
These results are obtained by solving the \emph{first-order model-checking problem} on those graph classes.
The input to this problem is a first-order sentence $\phi$ and a graph $G$
and the task is to decide whether $\phi$ is satisfied on $G$ or not.
This problem can trivially be solved in time $O(|G|^{|\phi|})$, which is optimal under complexity theoretic assumptions~\cite{chen2006strong}.
A first-order model-checking algorithm  is considered \emph{efficient} on some graph class if it solves the problem
in time $f(|\phi|)|G|^{O(1)}$ for some function $f$, i.e., in \emph{fpt time}.

\begin{figure}[t]
\begin{center}
\def\borderthickness{0.6}
\begin{tikzpicture}[align=center,scale=0.90] 
    \def\dstdown{1.6}
    \def\dstside{2.2}
    \tikzset{>=latex}
    \tikzstyle{edge}=[->, line width=\borderthickness]
    \tikzstyle{vertex1}=[rectangle, rounded corners, draw=black, minimum width=3.3cm, minimum height=1cm, line width=\borderthickness]
    \tikzstyle{vertex2}=[rectangle, rounded corners, draw=black, minimum width=1.5cm, minimum height=1cm, line width=\borderthickness]
    \node at (0,1.4) () {\bf Sparse}; 
    \node at (8,1.4) () {\bf Dense}; 
    \node[vertex1, fill = black!00]                                  at (0,0)             (nwd)  {Nowhere dense}; 
    \node[vertex1, fill = black!00]                                  at (0,-1*\dstdown)           (be)   {Bounded expansion}; 
    \node[vertex1, fill = black!00] at (8,0)             (snwd) {Structurally \\ nowhere dense}; 
    \node[vertex1, fill = black!00] at (8,-1*\dstdown)           (sbe)  {Structurally \\ bounded expansion}; 
    \node[vertex1, fill = black!00]                                  at (8,-3*\dstdown)           (sbd)  {Structurally \\ bounded degree }; 
    \node[vertex2, fill = black!00]                                  at (-1.3,-2*\dstdown)   (em)   {Excluded \\ minor}; 
    \node[vertex2, fill = black!00]                                  at (-\dstside,-3*\dstdown)   (p)    {Planar}; 
    \node[vertex2, fill = black!00]                                  at (0,-3*\dstdown)           (tw)   {Bounded \\ treewidth}; 
    \node[vertex2, fill = black!00]                                  at (\dstside,-3*\dstdown)    (bd)   {Bounded \\ degree}; 

    \node[]  at (11.5,0)                     (dummy) {~}; 

    \draw [dashed,draw=black!30, line width=\borderthickness] (4,1.6) -- (4,-3.6*\dstdown + 0.35);

    \draw[edge] (be) to node[above,fill=white]  {transduction} (sbe) {};
    \draw[edge] (nwd) to node[above,fill=white] {transduction} (snwd) {};
    \draw[edge] (bd) to node[above,pos=0.6] {transd.} (sbd) {};
    \draw[edge] (sbd) to (sbe) {};
    \draw[edge] (be) to (nwd) {};
    \draw[edge] (sbe) to (snwd) {};
    \draw[edge] (em) to (be) {};
    \draw[edge] (p) to (em) {};
    \draw[edge] (tw) to (em) {};
    \draw[edge] (bd) to (be) {};

    \tikzset{every loop/.style={min distance=10mm,in=-10,out=10,looseness=9}}
    \path[edge] (sbe) edge  [loop right] node  {\hspace{-1.4cm}transd.} ();
    \path[edge] (snwd) edge  [loop right] node {\hspace{-1.4cm}transd.} ();
    \path[edge] (sbd) edge  [loop right] node  {\hspace{-1.4cm}transd.} ();
\end{tikzpicture}
\end{center}
\vspace{-0.5cm}
\caption{Hierarchy of selected properties of sparse graph classes and transductions thereof. 
    For all of them, the first-order model-checking problem can be solved in fpt time.}
\label{figure}
\end{figure}

\Nesetril and Ossona de Mendez introduced \emph{bounded expansion} and \emph{nowhere dense} graph classes.
These are very robust notions of sparsity that generalize the previously mentioned graph classes (see left side of \Cref{figure})
and have many interesting algorithmic properties. 
Most notably, \Dvorak, \Kral and Thomas~\cite{dvorak2010deciding} solve the first-order model-checking problem in linear fpt time on bounded expansion classes,
and Grohe, Kreutzer and Siebertz~\cite{GroheKS2017} solve this problem in almost linear fpt time on nowhere dense graph classes.
For sparse graphs, this is in a sense the best possible result of this type:
If a graph class is monotone (i.e., closed under taking subgraphs),
then the model-checking problem is fpt if and only if the class is nowhere dense (under standard complexity assumptions).

We therefore have reached a natural barrier in the study of meta-theorems for sparse graphs.
But there are other well-structured graph classes
that do not fit into the framework of sparsity.
One of the current main goals in this area is to push the theory to account for \emph{dense} (or non-monotone) classes as well.
In particular, we want to find dense, but \emph{structurally simple} graph classes
on which one can solve the first-order model-checking problem in fpt time.

An established tool to capture such graph classes are \emph{first-order interpretations and transductions}.
For a given first-order formula $\phi(x,y)$, the corresponding (one dimensional\footnote{The common definition in model theory allows formulas \(\phi(\bar x, \bar y)\) with tuples \(\bar x\) and \(\bar y\) of free variables. We do not need this.}) \emph{first-order interpretation} 
is a function that maps an input a graph $G$ 
to the graph $I_\phi(G)$ with vertex set $V(G)$ and edge set $\{ uv \mid u,v \in V(G), u \neq v, G \models \phi(u,v) \lor \phi(v,u) \}$.
For example, with $\phi(x,y) = \neg \textit{edge}(x,y)$, the interpretation complements the input graph.
With $\phi(x,y) = \exists z \, \textit{edge}(x,z) \land \textit{edge}(z,y)$ it computes the square of a graph.
For a given formula $\phi(x,y)$, the \emph{first-order interpretation of a graph class $\G$}
is the graph class $\{ I_\phi(G) \mid G \in \G\}$.
\emph{First-order transductions} are slightly more powerful than interpretations, since they also have the ability to copy, delete and nondeterministically color vertices.
As the precise definition of transductions is not important for now, we delay it until \Cref{sec:prelim}.
We say a graph class $\G$ has \emph{structurally property $X$}
if it is the transduction of a graph class with property $X$ (see right side of \Cref{figure} for examples).
For the properties considered in this paper, this is equivalent to saying that $\G$ consists of induced subgraphs of an interpretation of a class with property $X$.
This definition has nice closure properties in the sense that applying a transduction
to a structurally property $X$ class yields again a structurally property $X$ class.

There have been recent efforts to lift the meta-theorems for sparse graph classes to transductions of these classes.
\Gajarsky et al.\ solve the first-order model-checking problem in fpt time for structurally bounded degree graph classes~\cite{gajarsky2016new}.
Bonnet et al.\ lift this to structurally bounded local cliquewidth~\cite{mccw}.
Very recently, tractability was shown for structurally nowhere dense classes~\cite{mcsnd}, generalizing the previous results.

Such algorithmic results build upon structural decompositions of the corresponding graph classes.
In this paper, we characterize structurally bounded expansion via
\emph{lacon-decompositions}, \emph{shrub-decompositions} and \emph{parity-decompositions}.
The techniques we introduce
were later used to provide \emph{quasi-bush decompositions}~\cite{dreier2022treelike} of similar flavor for structurally nowhere dense classes,
which were in turn an important ingredient for the aforementioned model-checking algorithm on these classes~\cite{mcsnd}.

Before, bounded expansion classes have been characterized via \emph{low shrubdepth covers},
where shrubdepth~\cite{de2019shrub,originalshrubdepth} can be understood as a dense equivalent to treedepth.
This nicely complements the characterization of bounded expansion via \emph{low treedepth covers}~\cite{nevsetvril2015low}.

We should mention another important approach to obtain dense, structurally simple graph classes.
Bonnet, Kim, Thomass{\'e} and Watrigant~\cite{twinwidth1} introduce a graph parameter called \emph{twinwidth}
that captures many interesting non-monotone graph classes. 
The authors provide an algorithm that --- given an appropriate twinwidth contraction sequence --- 
solves the first-order model-checking problem in linear fpt time~\cite{twinwidth1}.
Bounded twinwidth is orthogonal to structurally bounded expansion or structurally nowhere denseness.
The notion of \emph{monadically NIP} generalizes all aforementioned notions and is conjectured
to precisely capture the hereditary tractable graph classes.

\subsection{Introducing Lacon-, Shrub-  and Parity-Decompositions}

A lacon-, shrub-  or parity-decomposition of a graph $G$
is a labeled graph $G'$ that encodes $G$ in a possibly much sparser way.
These decompositions are not tied to any fixed sparsity measure,
instead one can characterize different dense graph classes by applying different sparsity requirements to $G'$.
In this paper, we use them to capture structurally bounded expansion, structurally bounded treewidth and structurally bounded treedepth, i.e., shrubdepth
(\autoref{thm:BEcharacterization}, \ref{thm:TDcharacterization}, \ref{thm:TWcharacterization}).
Due to their flexibility, it seems quite likely that they can be used to capture transductions of other sparse classes as well.
Generally speaking, we prove for certain properties \(X\) that a graph class has structurally property $X$ if and only if
each graph in the graph class has a lacon-, shrub-  or parity-decomposition with property $X$.

A decomposition of a graph $G$ consists of the so-called \emph{target vertices} $V(G)$ and additional \emph{hidden vertices}.
``Lacon'' is an acronym for \textbf{la}rgest \textbf{co}mmon \textbf{n}eighbor,
since there is an order on the hidden vertices and the label of the largest common neighbor of two vertices
(either ``0'' or ``1'')  determines whether they are adjacent or not.
Formally, this is defined as follows.
A linear order $\pi$ on the vertices of a graph $G$ is represented by an injective function $\pi \colon V(G) \to \N$.


\begin{defi}[Lacon-decomposition]\label{def:lacon}
    A \emph{lacon-decomposition} is a tuple $(L,\pi)$ satisfying the following properties.

    \begin{enumerate}
        \item $L$ is a bipartite graph with sides $T,H$.
                We say $T$ are the \emph{target} vertices and $H$ are the \emph{hidden} vertices of the decomposition.
        \item Every hidden vertex is labeled with either ``0'' or ``1''.

        \item $\pi$ is a linear order on the vertices of $L$ with $\pi(h) < \pi(t)$ for all $t \in T$, $h \in H$.

        \item For all $t,t' \in T$ holds $N(t) \cap N(t') \neq \emptyset$.
        The largest vertex with respect to $\pi$ in $N(t) \cap N(t')$
        is called the \emph{dominant vertex} between $t$ and $t'$.
    \end{enumerate}
    We say $(L,\pi)$ is a \emph{lacon-decomposition of a graph $G$} if
    \begin{enumerate}\setcounter{enumi}{4}
        \item $V(G) = T$,
        \item for all $t \neq t' \in T$ there is an edge between $t$ and $t'$ in $G$ if and only if
            the dominant vertex of $t$ and $t'$ is labeled with ``1''.
    \end{enumerate}
\end{defi}

\noindent This definition leads to an illustrative process that reconstructs a graph $G$ from its lacon-decomposition $(L,\pi)$:
In the beginning, all edges in $G$ are unspecified.
Then we reveal the hidden vertices of $L$ one by one in ascending order by $\pi$.
If we encounter a hidden vertex $h$ with label ``1'' we fully connect its neighborhood $N(h)$ in $G$.
On the other hand, if $h$ has label ``0'' we remove all edges between vertices from $N(h)$ in $G$.
The edges inserted or removed by $h$ ``overwrite'' the edges that were inserted or removed by a previous hidden vertex.
After all hidden vertices have been revealed, the resulting graph is exactly $G$.
See the middle \autoref{exampledecompositions} for an example of a lacon-decomposition.

Before we use lacon-decompositions to characterize transductions of different sparse graph classes,
let us also introduce shrub-decompositions.
Our definition closely follows the wording used by Ganian et al.~\cite{originalshrubdepth,de2019shrub} to define shrubdepth.

\begin{defi}[Shrub-decomposition] \label{def:shrub} 
Let $m$ and $d$ be nonnegative integers. 
A \emph{shrub-decomposition of $m$ colors and diameter $d$} of a graph $G$ is a pair
$(F,S)$ consisting of an undirected graph $F$ and a set
$S\subseteq\{1,2,\ldots,m\}^2\times \{1,2,\ldots,d\}$ (called \emph{signature}) such that 
\begin{enumerate}
    \item the distance in $F$ between any two vertices is at most $d$,
    \item the set of pendant vertices (i.e., degree-one vertices) of $F$ is exactly the set $V(G)$ of vertices of $G$,
    \item each pendant vertex of $F$ is assigned one of the colors $\{1,2\dots,m\}$, and
    \item \label{item:shrub-decomposition-edge} 
        for any $i,j,l$ it holds $(i,j,l)\in S$ iff $(j,i,l)\in S$
        (symmetry in the colors),
        and for any two vertices $u,v\in V(G)$ such that $u$ is colored with $i$ and $v$
        is colored with $j$ and the distance between $u,v$ in $F$ is $l$,
        the edge $uv$ exists in $G$ if and only if $(i,j,l)\in S$.
\end{enumerate}
\end{defi}

\noindent Item \myitem{shrub-decomposition-edge} says that the existence of a $G$-edge between $u,v\in V(G)$ depends only on the colors of $u,v$ and
their distance in the decomposition.
See the right side of \Cref{exampledecompositions} for an example.
A shrub-decomposition is a generalization of a \emph{tree-model} used in the definition of shrubdepth.
In fact, we can define shrubdepth (i.e., structurally bounded treedepth) using shrub-decompositions.
\begin{defi}[Shrubdepth~\cite{de2019shrub,originalshrubdepth}]
    We say a \emph{tree-model of $m$ colors and depth $d$}
    is a shrub-decomposition $(F,S)$ of $m$ colors where $F$ is a rooted
    tree and the length of every root-to-leaf path in $F$ is exactly $d$.
    A graph class $\G$ is said to have \emph{bounded shrubdepth} 
    if there exist numbers $m$ and $d$ such that every graph in $\G$ has a tree-model of $m$ colors and depth~$d$.
\end{defi}

Instead of requiring $F$ to be a tree, we can require $F$ to come from a bounded expansion class
to obtain a characterization of structurally bounded expansion, as we will see soon.
At last, we also introduce parity-decompositions, where adjacencies between vertices
is encoded by the parity of the intersection of their neighborhoods in the decomposition.

\begin{defi}[Parity-Decomposition]\label{def:parity}
    Let $d$ be a nonnegative integer. 
    A \emph{parity-decom\-po\-sition of target-degree $d$} is a bipartite graph $P$ with sides $T,H$, where every vertex in $T$ has degree at most $d$.
    We say $T$ are the \emph{target} vertices and $H$ are the \emph{hidden} vertices of the decomposition.
    We say $P$ is a \emph{parity-decomposition of a graph $G$} if
    $V(G) = T$, and for all $t \neq t' \in T$ there is an edge between $t$ and $t'$ in $G$ if and only if $|N(t) \cap N(t')|$ is odd.
\end{defi}

\begin{figure}
\begin{center}

\def\ty{-1.2}
\def\tdy{1.2}
\def\tdx{2}
\def\ybla{0}

\vspace{-1cm}

\scalebox{0.5}{
\begin{tikzpicture}[align=center,scale=1.5] 
    \def\borderthickness{1}
    \tikzstyle{target}=[circle,draw=black,fill=gray!40, minimum width=0.7cm]
    \tikzstyle{targetedge}=[draw=black,line width=\borderthickness]

    \path (0,0) rectangle (0,0.3);

    \node[target] at (0-1*\tdx,\ty-0*\tdy)                     (a) {}; 
    \node[target] at (0-1*\tdx,\ty-1*\tdy)                     (b) {}; 
    \node[target] at (0-0*\tdx,\ty-0*\tdy)                     (c) {}; 
    \node[target] at (0-0*\tdx,\ty-1*\tdy)                     (d) {}; 
    \node[target] at (0+1*\tdx,\ty-0.5*\tdy)                   (e) {}; 

    \draw[targetedge] (a) to (b) {};
    \draw[targetedge] (c) to (d) {};
    \draw[targetedge] (a) to (c) {};
    \draw[targetedge] (b) to (d) {};
    \draw[targetedge] (c) to (e) {};
    \draw[targetedge] (d) to (e) {};

\end{tikzpicture}
}
\bigskip
\medskip

\scalebox{0.5}{
\begin{tikzpicture}[align=center,scale=1.5] 
    \def\borderthickness{1}
    \tikzstyle{circle0}=[draw=black, minimum height=0.6cm,minimum width=0.6cm]
    \tikzstyle{circle1}=[text=white,fill=black, minimum height=0.6cm,minimum width=0.6cm]
    \tikzstyle{target}=[circle,draw=black,fill=gray!40, minimum width=0.7cm]
    \tikzstyle{edge1}=[draw=black, line width=\borderthickness]
    \tikzstyle{edge0}=[draw=black,dotted,  line width=\borderthickness]
    \tikzstyle{targetedge}=[draw=black!10,line width=3*\borderthickness]

    \def\xbla{1}

    \node[circle1] at (-3*\xbla,\ybla)    (1) {\Large 1}; 
    \node[circle0] at (-1*\xbla,\ybla)    (2) {\Large 0}; 
    \node[circle0] at (1*\xbla, \ybla)    (3) {\Large 0}; 
    \node[circle1] at (3*\xbla, \ybla)    (4) {\Large 1}; 

    \node[target] at (0-1*\tdx,\ty-0*\tdy)                     (a) {}; 
    \node[target] at (0-1*\tdx,\ty-1*\tdy)                     (b) {}; 
    \node[target] at (0-0*\tdx,\ty-0*\tdy)                     (c) {}; 
    \node[target] at (0-0*\tdx,\ty-1*\tdy)                     (d) {}; 
    \node[target] at (0+1*\tdx,\ty-0.5*\tdy)                   (e) {}; 

    \draw[edge1] (1) to (a) {};
    \draw[edge1] (1) to (b) {};
    \draw[edge1] (1) to (c) {};
    \draw[edge1] (1) to (d) {};

    \draw[edge0] (3) to (a) {};
    \draw[edge0] (3) to (d) {};
    \draw[edge0] (3) to (e) {};

    \draw[edge0] (2) to (b) {};
    \draw[edge0] (2) to (c) {};
    \draw[edge0] (2) to (e) {};

    \draw[edge1] (4) to (c) {};
    \draw[edge1] (4) to (d) {};
    \draw[edge1] (4) to (e) {};
\end{tikzpicture}
}~~~~~~~%
\scalebox{0.5}{
\begin{tikzpicture}[align=center,scale=1.5] 
    \def\borderthickness{1}
    \tikzstyle{hidden}=[draw=black,fill=gray!40, minimum height=0.6cm, minimum width=0.6cm]
    \tikzstyle{target0}=[circle,draw=black,fill=black, minimum width=0.7cm]
    \tikzstyle{target1}=[circle,draw=black,fill=white, minimum width=0.7cm]
    \tikzstyle{edge1}=[draw=black, line width=\borderthickness]
    \tikzstyle{edge0}=[draw=black,dotted,  line width=\borderthickness]
    \tikzstyle{targetedge}=[draw=black!10,line width=3*\borderthickness]

    \def\xbla{0.35}

    \path (0,0) rectangle (0,0.3);

    \node[hidden] at (-1,\ybla-0.5)      (1) {}; 
    \node[hidden] at (-1,\ybla+0.5)      (2) {}; 
    \node[hidden] at (1,       \ybla)      (3) {}; 
    \draw[edge1]              (1) to (3) {};
    \draw[edge1]              (1) to (2) {};
    \draw[edge1]              (2) to (3) {};

    \node[target1] at (0-1*\tdx,\ty-0*\tdy)                     (a) {}; 
    \node[target1] at (0-1*\tdx,\ty-1*\tdy)                     (b) {}; 
    \node[target0] at (0-0*\tdx,\ty-0*\tdy)                     (c) {}; 
    \node[target0] at (0-0*\tdx,\ty-1*\tdy)                     (d) {}; 
    \node[target0] at (0+1*\tdx,\ty-0.5*\tdy)                   (e) {}; 

    \draw[edge1] (2) to (a) {};
    \draw[edge1] (2) to (c) {};
    \draw[edge1] (1) to (b) {};
    \draw[edge1] (1) to (d) {};
    \draw[edge1] (3) to (e) {};

\end{tikzpicture}
}~~~~~~~%
\scalebox{0.5}{
\begin{tikzpicture}[align=center,scale=1.5] 
    \def\borderthickness{1}
    \tikzstyle{hidden}=[draw=black,fill=gray!40, minimum height=0.6cm, minimum width=0.6cm]
    \tikzstyle{target}=[circle,draw=black,fill=gray!40, minimum width=0.7cm]
    \tikzstyle{edge1}=[draw=black, line width=\borderthickness]
    \tikzstyle{edge0}=[draw=black,dotted,  line width=\borderthickness]
    \tikzstyle{targetedge}=[draw=black!10,line width=3*\borderthickness]

    \def\xbla{1}

    \node[hidden] at (-3*\xbla,\ybla)    (1) {}; 
    \node[hidden] at (-1*\xbla,\ybla)    (2) {}; 
    \node[hidden] at (1*\xbla, \ybla)    (3) {}; 
    \node[hidden] at (3*\xbla, \ybla)    (4) {}; 

    \node[target] at (0-1*\tdx,\ty-0*\tdy)                     (a) {}; 
    \node[target] at (0-1*\tdx,\ty-1*\tdy)                     (b) {}; 
    \node[target] at (0-0*\tdx,\ty-0*\tdy)                     (c) {}; 
    \node[target] at (0-0*\tdx,\ty-1*\tdy)                     (d) {}; 
    \node[target] at (0+1*\tdx,\ty-0.5*\tdy)                   (e) {}; 

    \draw[edge1] (1) to (a) {};
    \draw[edge1] (1) to (b) {};

    \draw[edge1] (2) to (b) {};
    \draw[edge1] (2) to (d) {};

    \draw[edge1] (3) to (a) {};
    \draw[edge1] (3) to (c) {};

    \draw[edge1] (4) to (c) {};
    \draw[edge1] (4) to (d) {};
    \draw[edge1] (4) to (e) {};
\end{tikzpicture}
}~~~~~~~%
\caption{%
    Top: A graph $G$.
    Left: A lacon-decomposition of $G$. The hidden vertices are listed in ascending order from left to right.
    Middle: A shrub-decomposition of $G$ with two colors and diameter three.
    Vertices from $G$ are adjacent if they have distance two or distance three and the same color.
    Right: A parity-decomposition of $G$ with target-degree two.
}
\label{exampledecompositions}
\end{center}
\end{figure}

\subsection{Generalized Coloring Numbers and Bounded Expansion}\label{sec:coloringnumbers}

Before we present our results, we introduce the so-called
\emph{generalized coloring numbers}, as we use them to pose a sparsity requirement on lacon-decompositions.
These numbers can be used to characterize bounded expansion, treewidth and treedepth
and have numerous algorithmic applications. 
Let $G$ be an undirected graph and $\pi$ be an ordering on the vertices of $G$.
We say a vertex $u$ is \emph{$r$-reachable} from $v$ with respect to $\pi$ if $\pi(u) \le \pi(v)$ and there is a path of length at most $r$
from $v$ to $u$ and for all vertices $w$ on the path either $w = u$ or $\pi(w) \ge \pi(v)$.
A vertex $u$ is \emph{weakly $r$-reachable} from $v$ with respect to $\pi$ if there is a path of length at most $r$
from $v$ to $u$ and $\pi(u) \le \pi(w)$ for all vertices $w$ on that path.
If $G$ is a directed graph, we say $u$ is (weakly) $r$-reachable from $v$ if and only if it is (weakly) $r$-reachable in the underlying undirected graph.

Let $\reach_r(G,\pi,v)$ be the set of vertices that are $r$-reachable from $v$ with respect to $\pi$.
We similarly define $\wreach_r(G,\pi,v)$ be the set of weakly $r$-reachable vertices.
We~set
\[
\col_r(G,\pi) = \max_{v \in V(G)} |\reach_r(G,\pi,v)|,
\]
\[
\wcol_r(G,\pi) = \max_{v \in V(G)} |\wreach_r(G,\pi,v)|.
\]
\goodbreak
We define $\Pi(G)$ to be the set of all linear orders on $G$.
Finally, the \emph{$r$-coloring number} and \emph{weak $r$-coloring number} of a graph $G$ is defined as
\[
\col_r(G) = \min_{\pi \in \Pi(G)} \col_r(G,\pi),
\]
\[
\wcol_r(G) = \min_{\pi \in \Pi(G)} \wcol_r(G,\pi).
\]

For small $r$, the two flavors of generalized coloring numbers are strongly related.
It holds that $\col_r(G) \le \wcol_r(G) \le \col_r(G)^r$~\cite{coloringdefinition}.
For large $r$, the generalized coloring numbers converge to treewidth ($\tw(G)$) and treedepth ($\td(G)$), respectively~\cite{NesetrilM2012,coloringcovering}
\[
\col_1(G) \le \dots \le \col_\infty(G) = \tw(G)+1,
\]
\[
\wcol_1(G) \le \dots \le \wcol_\infty(G) = \td(G).
\]

A graph $H$ is an \emph{$r$-shallow minor} of a graph $G$
if it is the result of first contracting mutually disjoint connected subgraphs with radius at most $r$ in $G$ and then taking a subgraph.
A graph class $\G$ has \emph{bounded expansion} if there exists a function $f(r)$ such that $\frac{|E(H)|}{|V(H)|} \le f(r)$
for all $G \in \G$ and $r$-shallow minors $H$ of~$G$~\cite{NesetrilM2012}.
Zhu first observed that generalized coloring numbers can also be used to characterize bounded expansion~\cite{zhu2009colouring}.
In this paper, we rely heavily on the following characterization by van den Heuvel and Kierstead~\cite{van2019uniform},
which is slightly stronger than Zhu's~\cite{zhu2009colouring} original characterization.
\begin{defi}[Bounded Expansion~\cite{van2019uniform}]\label{def:be}
    A graph class $\G$ has \emph{bounded expansion} if there
    exists a function $f(r)$ such that every graph $G \in \G$ has an ordering $\pi$ of its vertices 
    with $\col_r(G,\pi) \le f(r)$ for all $r$.
\end{defi}
As illustrated in~\Cref{figure}, examples of bounded expansion graph classes include 
planar graphs or any graph class of bounded degree.

\subsection{Main Result}

This paper presents characterizations of structurally
bounded expansion based on lacon-, shrub-  and parity-decompositions.
As a side result, we also obtain characterizations of structurally bounded treewidth and treedepth.
For the definition of \emph{first-order transductions} and \emph{structurally bounded expansion} see~\Cref{sec:prelim}.
The following theorem contains the main contribution of this paper.

\begin{thm}\label{thm:BEcharacterization}
Let $\G$ be a graph class.
The following statements are equivalent.
\begin{enumerate}\setlength\itemsep{0.2em}
    \item\label{item:bebase}
        $\G$ has structurally bounded expansion, i.e.,
        there exists a graph class $\G'$ with bounded expansion and a first-order transduction $\tau$ such that $\G \subseteq \tau(\G')$.
    \item\label{item:belacon}
        There exists a function $f(r)$ such that every graph in $\G$
        has a lacon-decomposition $(L,\pi)$ with $\col_r(L,\pi) \le f(r)$ for all $r$
        (this implies that $L$ comes from a bounded expansion class).
    \item\label{item:beshrub}
        There exist a signature $S$, a number $d$ and a graph class $\G'$ with bounded expansion such that
        every graph in $\G$ has a shrub-decomposition $(F,S)$ with one color, diameter at most~$d$ and $F \in \G'$.
    \item\label{item:beparity}
        There exist number $d$ and a graph class $\G'$ with bounded expansion such that
        every graph in $\G$ has a parity-decomposition $P \in \G'$ with target-degree at most~$d$.
\end{enumerate}
\end{thm}


Equivalently, structurally bounded expansion classes can also be characterized via
\emph{low shrubdepth covers}~\cite{gajarsky2018first}.
This characterization was obtained using quantifier elimination, which is possible on bounded expansion~\cite{DvorakKT2013}, but not on nowhere dense classes~\cite{no-qe}.
The techniques of this paper (local separators and logical composition theorems) have a better chance to generalize to structurally nowhere dense classes and beyond.
In fact, in a follow-up paper~\cite{dreier2022treelike}, we build upon this approach to derive shrub-decompositions as well as low shrubdepth covers for structurally nowhere dense graph classes.

\autoref{thm:BEcharacterization} can further be understood as a limit on the expressive power of transductions on bounded expansion classes.
For example, it implies that transductions based on boolean combinations of local, purely existential formulas have the same expressive power as general first-order transductions:
Assume a graph class $\cal G$ was obtained from a graph class $\cal G'$ with bounded expansion via a (possibly very complicated) transduction.
We replace $\cal G'$ with the class of shrub-decompositions of $\cal G$, as described in \myitem{beshrub}.
Now $\cal G$ can be expressed as a very simple transduction of $\cal G'$.
We merely have to check the color of two vertices and their (bounded) distance.
This can be done using a boolean combination of local, existential formulas.
When introducing function symbols \(f_i(v)\) for the \(i\)th hidden-neighbor of a target vertex \(v\) in a lacon-decomposition,
this further characterizes structurally bounded expansion classes as (functional) quantifier-free transductions of bounded expansion classes.

Other bounds on the expressive power of transductions on sparse graphs (orthogonal to ours) have been obtained in the literature~\cite{gajarsky2018first,nesetril2020towards}.
As side results, we obtain the following characterizations of structurally bounded treedepth and treewidth.
We should note that structurally bounded treedepth is the same as bounded shrubdepth~\cite{originalshrubdepth}.

\begin{thm}\label{thm:TDcharacterization}
Let $\G$ be a graph class.
The following statements are equivalent.
\begin{enumerate}\setlength\itemsep{0.2em}
    \item\label{item:tdbase}
        $\G$ has structurally bounded treedepth, i.e.,
        there exists a graph class $\G'$ with bounded treedepth and a first-order transduction $\tau$ such that $\G \subseteq \tau(\G')$.
    \item\label{item:tdlacon}
        There exists a number $d$ such that every graph in $\G$
        has a lacon-decomposition $(L,\pi)$ with $\wcol_\infty(L,\pi) \le d$
        (this implies $L$ has treedepth at most $d$).
    \item\label{item:tdshrub}
        There exist a number $d$ and a signature $S$ such that every graph in $\G$
        has a shrub-decomposition $(F,S)$ with one color, diameter at most~$d$ and treedepth at most~$d$.
    \item\label{item:tdparity}
        There exist number $d$ and a graph class $\G'$ with bounded treedepth such that
        every graph in $\G$ has a parity-decomposition $P \in \G'$ with target-degree at most~$d$.
\end{enumerate}
\end{thm}

\begin{thm}\label{thm:TWcharacterization}
Let $\G$ be a graph class.
The following statements are equivalent.
\begin{enumerate}
    \item\label{item:twbase}
        $\G$ has structurally bounded treewidth, i.e.,
        there exists a graph class $\G'$ with bounded treewidth and a first-order transduction $\tau$ such that $\G \subseteq \tau(\G')$.
    \item\label{item:twlacon}
        There exists a number $t$ such that every graph in $\G$
        has a lacon-decomposition $(L,\pi)$ with $\col_\infty(L,\pi) \le t-1$
        (this implies $L$ has treewidth at most $t$).
    \item\label{item:twshrub}
        There exist a number $t$ and a signature $S$ such that every graph in $\G$
        has a shrub-decomposition $(F,S)$ with one color, diameter at most~$t$ and treewidth at most~$t$.
    \item\label{item:twparity}
        There exist number $d$ and a graph class $\G'$ with bounded treewidth such that
        every graph in $\G$ has a parity-decomposition $P \in \G'$ with target-degree at most~$d$.
\end{enumerate}
\end{thm}

The results of \autoref{thm:BEcharacterization}, \ref{thm:TDcharacterization}, and \ref{thm:TWcharacterization}
are consequences of a more general statement, \autoref{thm:maindirection},
which shows that for every transduction of a graph $G$, we can find
an equivalent lacon-decomposition whose generalized coloring numbers are not too far off from the numbers of $G$.

\subsection{Localized Feferman--Vaught Composition Theorem}

The Feferman--Vaught theorem~\cite{FV59} states that the validity of first-order formulas
on the disjoint union or Cartesian product of two graphs is uniquely
determined by the value of first-order formulas on the individual graphs.
Makowsky adjusted the theorem for algorithmic
use in the context of MSO model-checking~\cite{makowsky2004algorithmic}. 
It has numerous applications and in the area of meta-theorems it leads, for example,
to an especially concise proof of Courcelle's theorem~\cite{grohe2008logic}.

A highly useful property of first-order logic is \emph{locality},
for example, in the form of Hanf's~\cite{Hanf1965} or Gaifman's~\cite{gaifman1982local} theorem.
Intuitively, these theorems state that first-order logic can only express local properties.
Locality is a key ingredient in many first-order model-checking algorithms~\cite{fomc,GroheKS2017,KuskeS2017,Seese1996,mccw,mcsnd}.

A central building block in our proofs is a localized variant of the Feferman--Vaught theorem (also proved in~\cite[Lemma 15]{pilipczuk2018number}).
The essence of this result is the following:
Assume we have a graph $G$, a first-order formula $\phi(x,y)$ and two vertices $v,w$ and
want to know whether $G \models \phi(v,w)$.
Further assume that we have some kind of ``local separator'' between $v,w$,
i.e., a tuple of vertices $\bu$ such that all short paths between $v$ and $w$
pass through $\bu$.
We show that knowing whether $G \models \psi(\bu,v)$, $G \models \psi(\bu,w)$
for certain formulas $\psi$ this gives us enough information to compute
whether $G \models \phi(v,w)$.
The original Feferman--Vaught theorem claims this only if $\bu$ is an actual separator between $v$ and $w$,
i.e., \emph{all} paths between $v$ and $w$ pass through $\bu$.
To formalize our result, we need to define so-called \emph{$q$-types}.
The \emph{quantifier rank} of a formula is the maximal nesting depth of its quantifiers.

\begin{restatable}[q-type~\cite{grohe2008logic}]{defi}{defqtype}\label{def:qtype}
    Let $G$ be a labeled graph and $\bar v = (v_1,\dots,v_k) \in V(G)^k$.
    The \emph{$q$-type of $\bar v$ in $G$} is
    the set $\tp(G,\bar v)$ of all first-order formulas $\psi(x_1\dots x_k)$ of quantifier rank at most $q$
    such that $G \models \psi(v_1\dots v_k)$.
\end{restatable}
\newcommand{\elements}{\textnormal{set}}
We syntactically normalize formulas so that there are
only finitely many formulas of fixed quantifier rank and
with a fixed set of free variables. 
Therefore $q$-types are finite sets.
For a tuple $\bar u = (u_1,\dots,u_k)$, we denote the set $\{u_1,\dots,u_k\}$ by $\elements(\bar u)$.
We follow Grohe's presentation of the Feferman--Vaught theorem~\cite{grohe2008logic}.
\begin{propC}[{\cite[Lemma~2.3]{grohe2008logic}}]\label{thm:fv}
    Let $G,H$ be labeled graphs and 
    $\bar u \in V(G)^k$,
    $\bar v \in V(G)^l$,
    $\bar w \in V(H)^m$,
    such that
    $V(G) \cap V(H) = \elements(\bu)$. 
    Then for all $q \in \N$, $\tp(G \cup H, \bar u \bar v \bar w)$
    is determined by $\tp(G, \bar u \bar v)$ and $\tp(H, \bar u \bar w)$.
\end{propC}

In the previous statement, $\bu$ separates $\bar v$ and $\bar w$ by splitting $G \cup H$ into the subgraphs $G$ and $H$.
We extend this result using a more general notion of separation that goes as follows.
The \emph{length} of a path equals its number of edges.

\begin{restatable}{defi}{defseparator}\label{def:separator}
Let $G$ be a graph, $r \in \N$, and $\bu$, $\bv_1, \dots, \bv_k$ be tuples of vertices from $G$.
We say $\bu$ \emph{$r$-separates} $\bv_i$ and $\bv_j$ if every path of length at most $r$ between a vertex from $\bv_i$ and a vertex from $\bv_j$ contains at least one vertex from $\bu$.
We say $\bu$ \emph{$r$-separates} $\bv_1, \dots, \bv_k$ if it $r$-separates $\bv_i$ and $\bv_j$ for all $i\neq j$.
In particular, if a vertex \(v\) appears in multiple tuples among \(\bv_1,\dots,\bv_k\), then an \(r\)-separating tuple \(\bu\) also needs to contain \(v\).
\end{restatable}

\noindent Based on this notion of separation,
we use the following Feferman--Vaught-inspired result.

\begin{restatable}{thm}{thmlocalfv}\label{thm:localfv}
    There exists a function $f(q,l)$ such that for all
    labeled graphs $G$, every $q,l \in \N$,
    and all tuples $\bu$, $\bv_1, \dots, \bv_k$ of vertices from $G$ such that $\bu$ $4^q$-separates $\bv_1,\dots,\bv_k$
    and $|\bu|+|\bv_1|+\dots+|\bv_k| \le l$,
    the type $\tp(G,\bu \bv_1  \dots \bv_k)$ depends only on the types $\tpp{f(q,l)}(G,\bu \bv_1), \dots, \tpp{f(q,l)}(G,\bu \bv_k)$.
    Furthermore, $\tp(G,\bu \bv_1 \dots \bv_k)$ can be computed from $\tpp{f(q,l)}(G,\bu \bv_1), \dots, \tpp{f(q,l)}(G,\bu \bv_k)$.
\end{restatable}
\goodbreak

\subsection{Techniques and Outline}

The two cornerstones of our proofs
are a localized Feferman--Vaught composition theorem~\cite{pilipczuk2018number}, as well as separators derived via generalized coloring numbers.
We show the following circular sequence of implications.
\begin{itemize}[leftmargin=2.3cm]
    \item[(i)]   
        We start with a graph $G$ that is a transduction of a graph $G'$, where $G'$ has bounded generalized coloring numbers.
    \item[(i) $\Longrightarrow$ (ii)]\label{itoii}
        Then we construct a so-called \emph{directed lacon-decomposition} $(L,\pi)$ of $G$ whose coloring numbers are bounded as well.
        This is a generalization of a lacon-decomposition where we allow $L$ to be a directed graph.
        This step forms the central part of the paper and here we use the localized Feferman--Vaught theorem extensively.
   \item[(ii) $\Longrightarrow$ (iii)]
       Next, we convert $(L,\pi)$ into a normal lacon-decomposition without increasing the generalized coloring numbers too much.
   \item[(iii) $\Longrightarrow$ (iv)]
        We transform the lacon-decomposition into an equivalent shrub-decompo\-sition $(F,S)$, also with bounded coloring numbers.
   \item[(iii) $\Longrightarrow$ (v)]
        We transform the lacon-decomposition into an equivalent parity-decompo\-sition $P$, also with bounded coloring numbers.
   \item[(iv,v) $\Longrightarrow$ (i)]
        This implies that $G$ is a transduction of $F$ or $P$, and since the generalized coloring numbers of both are bounded, this brings us back to the start.
\end{itemize}

Afterwards, we use the fact that generalized coloring numbers can describe bounded expansion, bounded treewidth and bounded treedepth.
Therefore, \autoref{thm:BEcharacterization}, \ref{thm:TDcharacterization} and \ref{thm:TWcharacterization}
follow by posing different bounds on the coloring numbers.

The proof of (i) $\Longrightarrow$ (ii)
is the central technical contribution of this paper.
We sketch some its ideas.
We have a graph $G'$ with an ordering $\pi$ of its vertices
and want to obtain a lacon-decomposition of an interpretation $I_\phi(G')$ of $G'$.
Consider vertices $v,w \in V(G')$ and a tuple $\bu$ consisting of the vertices in $\wreach_{r}(G',\pi,v) \cap \wreach_{r}(G',\pi,w)$.
It is a basic property of generalized coloring numbers that every path of length at most $r$ between $v$ and $w$ passes through $\bu$,
i.e., $\bu$ \emph{$r$-separates} $v$ and $w$.
The localized decomposition theorem states that $\phi(v,w)$ depends only on $q$-types of $\bu v$ and $\bu w$.
For every possible tuple $\bu$ and every combination of $q$-types, we introduce a hidden vertex that we label with ``1''
if and only if its two $q$-types together imply $\phi$ to be true.
We then connect the hidden vertices with the vertices of $G'$ if their $q$-types match.

The implication (i) $\Longrightarrow$ (ii) 
is proved in \Cref{sec:betodirlacon}
and (ii) $\Longrightarrow$ (iii) is proved in \Cref{sec:dirtoundir}.
This is then combined in \Cref{sec:onetothree} to show
(i) $\Longrightarrow$ (iii).
In \Cref{sec:lacontoshrub} we prove (iii) $\Longrightarrow$ (iv)
and in \Cref{sec:lacontoparity} we prove (iii) $\Longrightarrow$ (v).
Then, in \Cref{sec:threetheorems} we combine all these implications to prove our
main results \autoref{thm:BEcharacterization}, \ref{thm:TDcharacterization} and \ref{thm:TWcharacterization}.
The localized Feferman--Vaught theorem has been proved before~\cite[Lemma 15]{pilipczuk2018number}.
We nevertheless finish the paper in \Cref{sec:localfv} with a self-contained proof of the theorem.

\section{Preliminaries}\label{sec:prelim}

Any unusual graph notation is limited to specific sections and therefore introduced where it is used.
We use first-order logic over labeled and unlabeled graphs.
An unlabeled graph is a relational structure with a binary edge relation.
A labeled graph may additionally have labels (or colors) on vertices or edges, represented
by additional unary and binary relations.
The length of a formula \(\phi\) is denoted by \(|\phi|\).
For a formula \(\phi(x,y)\) and graph $G$, the \emph{interpretation} of $G$ under $\phi$, denoted by $I_\phi(G)$,
is the undirected unlabeled graph with vertex set $V(G)$ and edge set $\{ uv \mid u,v \in V(G), u \neq v, G \models \phi(u,v) \lor \phi(v,u) \}$.

\paragraph{Transductions.}
\emph{First-order transductions} extend interpretations with the ability to delete vertices, as well as copy and color the input graph.
We use the same notation as~\cite{gajarsky2016new}.
The central building block are \emph{basic transductions.}
A basic transduction is a triple $\tau_0 = (\chi, \nu, \phi)$ of first-order formulas of arity zero, one and two.
If $G \not\models \chi$, then $\tau_0(G)$ is undefined.
Otherwise $\tau_0(G)$ is the graph with vertex set \mbox{$\{ v \mid v \in V(G), G \models \nu(v)\}$}
and edge set $\{ uv \mid u,v \in V(G), u \neq v, G \models \phi(u,v) \lor \phi(v,u) \}$.
There are two more building blocks, which are less important for this work.
A \emph{$p$-parameter expansion} is an operation that maps each graph $G$ to the set of all graphs that can be created by
adding $p$ unary predicates (i.e., colors or labels) to $G$.
An \emph{$m$-copy operation} maps a graph $G$ to a graph $G^m$ with
$V(G^m) = V(G) \times \{1,\dots,m\}$ and
${E(G^m) = \{ (v,i)(w,i) \mid vw \in E(G), 1 \le i \le m\}} \cup \{ (v,i)(v,j) \mid v \in V(G), 1 \le i < j \le m \}$.
Furthermore, $G^m$ has a binary relation~$\sim$
labeling the edges $\{ (v,i)(v,j) \mid v \in V(G), 1 \le i < j \le m \}$
and unary relations $Q_1,\dots,Q_m$ with $Q_i = \{(v,i) \mid v~\in~V(G)\}$.
Thus $G^m$ consists of $m$ copies of $G$ with an equivalence relation \(\sim\) between copied vertices.
Finally, a \emph{transduction} $\tau$ is an operation of the form $\tau = \tau_0 \circ \gamma \circ \epsilon$
where $\epsilon$ is a $p$-parameter expansion,
$\gamma$ is a $m$-copy operation and 
$\tau_0$ is a basic transduction.
Notice that the output $\tau(G)$ is a \emph{set of graphs} because of the parameter expansion.
For a class $\cal G$, we define $\tau(\cal G) = \bigcup_{G \in \cal G} \tau(G)$ to be the \emph{transduction of $\cal G$}.
This now gives us the definition of structurally bounded expansion and related graph classes.
\begin{defi}[Structurally bounded expansion~\cite{gajarsky2018first}]\label{def:strucboundexp}
    A graph class $\G$ has \emph{structurally bounded expansion} if there
    exists a graph class $\G'$ with bounded expansion and a transduction $\tau$ such that $\G \subseteq \tau(\G')$.
    Generally speaking, $\G$ has \emph{structurally property $X$} if $\G'$ has property~$X$.
\end{defi}

\section{Interpretations of Bounded Expansion Have Directed Lacon-Decompositions}\label{sec:betodirlacon}

This section contains the central idea of this paper.
We consider a generalization of lacon-decompositions called \emph{directed lacon-decompositions}
and construct such a decomposition with bounded generalized coloring numbers.
In the following definition we denote the in- and out-neighborhoods of a vertex $h$ by $N^-(h)$ and $N^+(h)$, respectively.

\begin{defi}[Directed Lacon-decomposition]\label{def:dirlacon}
    A \emph{directed lacon-decomposition} is a tuple $(L,\pi)$ satisfying the following properties.
    \begin{enumerate}
        \item $L$ is a directed bipartite graph with sides $T,H$.
                We say $T$ are the \emph{target} vertices and $H$ are the \emph{hidden} vertices of the decomposition.
        \item Every hidden vertex is labeled with either ``0'' or ``1''.

        \item $\pi$ is a linear order on the vertices of $L$ with $\pi(h) < \pi(t)$ for all $t \in T$, $h \in H$.

        \item For all $t \neq t' \in T$ holds $\bigl(N^-(t) \cap N^+(t') \bigr) \cup \bigl(N^+(t) \cap N^-(t')\bigr) \neq \emptyset$.
        The largest vertex with respect to $\pi$ in $\bigl(N^-(t) \cap N^+(t') \bigr) \cup \bigl(N^+(t) \cap N^-(t')\bigr)$
        is called the \emph{dominant vertex} between $t$ and $t'$.
    \end{enumerate}
    We say $(L,\pi)$ is the \emph{directed lacon-decomposition of an undirected graph $G$} if
    \begin{enumerate}\setcounter{enumi}{4}
        \item $V(G) = T$,
        \item for all $t \neq t' \in T$ there is an edge between $t$ and $t'$ in $G$ if and only if
            the dominant vertex of $t$ and $t'$ is labeled with ``1''.
    \end{enumerate}
\end{defi}

We will need the following facts about generalized coloring numbers.
As they can be easily derived from their basic definition, we omit a proof.
\begin{prop}\label{prop:colfacts}
    Let $G$ be a graph with ordering $\sigma$ and $r \in \N$.
    \begin{enumerate}
        \item\label{item:2r}
        For $v \in V(G)$ and $u,w \in \wreach_{r}(G,\sigma,v)$ with $\sigma(w) \le \sigma(u)$ holds \\ $w \in \wreach_{2r}(G,\sigma,u)$.
    \item\label{item:rpath}
        Let $v,w \in V(G)$ and $S = \wreach_{r}(G,\sigma,v) \cap \wreach_{r}(G,\sigma,w)$.
        Then every path of length at most $r$ between $v$ and $w$ contains a vertex from $S$.
        Note that it is possible that \(v \in S\) or \(w \in S\), in which case the statement holds trivially.
    \item\label{item:bugfix-kr}
        Let $v_1,\dots,v_k$ be vertices with $\sigma(v_1) \le \sigma(v_k)$ and $\min_{i=2}^k \sigma(v_i) = \sigma(v_k)$
        such that for all $1\le i < k$ either
        $v_i \in \wreach_r(G,\sigma,v_{i+1})$ or 
        $v_{i+1} \in \wreach_r(G,\sigma,v_{i})$.
        Then there exists $w \in \reach_{rk}(G,\sigma,v_k)$ with $v_1 \in \wreach_{r}(G,\sigma,w)$.
    \item\label{item:weak-kr}
        Let $v_1,\dots,v_k$ be vertices with $\sigma(v_1) = \min_{i=1}^k \sigma(v_i)$
        such that for all $1\le i < k$ either
        $v_i \in \wreach_r(G,\sigma,v_{i+1})$ or 
        $v_{i+1} \in \wreach_r(G,\sigma,v_{i})$.
        Then $v_1 \in \wreach_{rk}(G,\sigma,v_k)$.
    \item\label{item:colwcol}
        $\wcol_r(G,\sigma) \le \col_r(G,\sigma)^r$.
    \end{enumerate}
\end{prop}

\begin{lem}\label{lem:interpretationtolacon}
    Let $\phi(x,y)$ be a first-order formula.
    There exists a function $g$ such that
    for every labeled graph $G$ and ordering $\sigma$ on the vertices of $G$
    there exists a directed lacon-decomposition $(L,\pi)$ of $I_\phi(G)$ with
    \begin{itemize}
        \item $\col_r(L,\pi) \le g\bigl(|\phi| + \col_{2 \cdot 4^{|\phi|}}(G,\sigma)\bigr) \cdot \col_{4^{|\phi|} r}(G,\sigma)$ for all $r$,
        \item $\wcol_r(L,\pi) \le g\bigl(|\phi| + \col_{2 \cdot 4^{|\phi|}}(G,\sigma)\bigr) \cdot \wcol_{4^{|\phi|} r}(G,\sigma)$ for all $r$.
    \end{itemize}
\end{lem}
\begin{proof}
    For technical reasons, it is easier to prove the result of this lemma under the
    additional assumption that the input graph $G$ has an apex vertex.
    We do so first, and at the end of the proof we generalize the result also
    to graphs without an apex vertex.
    Our main proof is outlined as follows:
    We first construct a directed lacon-decomposition, then prove that the construction encodes the graph correctly
    and at last bound the coloring numbers.

\paragraph{Constructing a Lacon-Decomposition.}
    Let us fix a graph $G$ (with an apex vertex) and an ordering $\sigma$.
    Let $q = |\phi|$, $l=\wcol_{2 \cdot 4^{q}}(G,\sigma)+2$ and let $f(q,l)$ be the function from \autoref{thm:localfv}.
    We define the hidden vertices $H$ of $L$ to be all tuples
    $(u,\bu,\typi_1,\typi_2)$ such that
    \begin{itemize}
        \item $u \in V(G)$,
        \item $\bu$ is the tuple of all vertices in $\wreach_{2 \cdot 4^{q}}(G,\sigma,u)$ ordered in ascending order by~$\sigma$,
        \item $\typi_1$, $\typi_2$ are $f(q,l)$-types containing formulas with $|\bu|+1$ free variables.
    \end{itemize}

    Let us fix one such hidden vertex $(u,\bu,\typi_1,\typi_2)$.
    Consider vertices $v_1,v_2 \in V(G)$ that are $4^q$-separated by $\bu$ 
    and have types $\tpp{f(q,l)}(G,\bu v_1) = \typi_1$ and $\tpp{f(q,l)}(G,\bu v_2) = \typi_2$.
    The fact whether $G \models \phi(v_1,v_2) \lor \phi(v_2,v_1)$ is determined by the type $\tpp{q}(G,\bu v_1 v_2)$.
    And according to the localized Feferman--Vaught variant in \autoref{thm:localfv}, $\tpp{q}(G,\bu v_1 v_2)$ is in turn determined by $\typi_1$ and $\typi_2$.
    With this in mind, 
    we iterate over all hidden vertices $(u,\bu,\typi_1,\typi_2)$ and distinguish two cases:

    \begin{itemize}
        \item \emph{Case 1:}
        For all $v_1,v_2 \in V(G)$ that are $4^{q}$-separated by $\bu$ with 
        $\tpp{f(q,l)}(G,\bu v_1) = \typi_1$ and $\tpp{f(q,l)}(G,\bu v_2) = \typi_2$
        holds $G \models \phi(v_1,v_2) \lor \phi(v_2,v_1)$.
        We give $(u,\bu,\typi_1,\typi_2)$ in this case the label~``1''.

        \item \emph{Case 2:}
        For all $v_1,v_2 \in V(G)$ that are $4^{q}$-separated by $\bu$ with 
        $\tpp{f(q,l)}(G,\bu v_1) = \typi_1$ and $\tpp{f(q,l)}(G,\bu v_2) = \typi_2$
        holds $G \not\models \phi(v_1,v_2) \lor \phi(v_2,v_1)$.
        We give $(u,\bu,\typi_1,\typi_2)$ in this case the label~``0''.
\end{itemize}

    Now, every hidden vertex is labeled with either ``1'' or ``0''.
    We define the arc set of our directed lacon-decomposition as follows:
    Let $(u,\bu,\typi_1,\typi_2)$ be a hidden vertex.
    For every $v_1$ with $\tpp{f(q,l)}(G,\bu v_1) = \typi_1$ such that $u \in \wreach_{4^{q}}(G,\sigma,v_1)$
    we add an arc from $v_1$ to $(u,\bu,\typi_1,\typi_2)$.
    Similarly, for every $v_2$ with $\tpp{f(q,l)}(G,\bu v_2) = \typi_2$ such that $u \in \wreach_{4^{q}}(G,\sigma,v_2)$
    we add an arc from $(u,\bu,\typi_1,\typi_2)$ to $v_2$.

    As the last step of our construction, we fix an ordering $\pi$ on the vertices of $L$ such that 
    for all hidden vertices $(u,\bu,\typi_1,\typi_2), (u',\bu',\typi_1',\typi_2') \in H$ with
    $\sigma(u) < \sigma(u')$ holds $\pi((u,\bu,\typi_1,\typi_2)) < \pi((u',\bu',\typi_1',\typi_2'))$.
    We further require for every $h \in H$ and every $t \in V(G)$ that $\pi(h) < \pi(t)$.
    Such an ordering trivially exists.

\paragraph{Correctness of Construction.}
    At first, we need to show that $(L,\pi)$ is a directed lacon-decomposition as defined in \autoref{def:dirlacon}.
    One can easily verify that $L$ is in fact a directed bipartite graph, every hidden vertex is either labeled with ``1'' or ``0''
    and that $\pi$ is an ordering on the vertices of $L$ with $\pi(h) < \pi(t)$ for all $h \in H$, $t \in T=V(G)$.
    What is left to do, is fix some vertices $v_1, v_2 \in T$ and show
    that the set $\bigl(N^-(v_1) \cap N^+(v_2) \bigr) \cup \bigl(N^+(v_1) \cap N^-(v_2)\bigr)$ is non-empty.
    We define $R_1 = \wreach_{4^{q}}(G,\sigma,v_1)$ and $R_2 = \wreach_{4^{q}}(G,\sigma,v_2)$.
    Since $G$ has an apex vertex, there is a path of length at most two between $v_1$ and $v_2$.
    We can also generally assume that $4^{q} \ge 2$ and therefore, by \myitem{rpath} of \autoref{prop:colfacts}, $R_1 \cap R_2 \neq \emptyset$.
    We choose some vertex $u' \in R_1 \cap R_2$
    and consider the hidden vertex $h' = (u',\bu',\tpp{f(q,l)}(G,\bu'v_1),\tpp{f(q,l)}(G,\bu'v_2))$ 
    where $\bu'$ is the tuple of all vertices in $\wreach_{2 \cdot 4^{q}}(G,\sigma,u')$, ordered in ascending order by $\sigma$.
    We constructed $L$ such that 
    there is an arc from $v_1$ to $h'$ and an arc from $h'$ to $v_2$.
    We conclude that $\bigl(N^-(v_1) \cap N^+(v_2) \bigr) \cup \bigl(N^+(v_1) \cap N^-(v_2)\bigr)$ is non-empty
    and therefore that $(L,\pi)$ is a directed lacon-decomposition.

    Next, we show that $(L,\pi)$ also is a directed lacon-decomposition of $I_\phi(G)$.
    $L$ was constructed such that its target vertices are $T = V(G) = V(I_\phi(G))$.
    It remains to show that the dominant vertex of two arbitrary vertices $v_1,v_2 \in T$ is labeled with ``1'' if and only if $v_1v_2 \in E(I_\phi(G))$.
    Let $h$ be the dominant vertex of $v_1$ and $v_2$.
    By construction, $h$ is either of the form $h = (u,\bu,\tpp{f(q,l)}(G,\bu v_1),\tpp{f(q,l)}(G,\bu v_2))$ or $h = (u,\bu,\tpp{f(q,l)}(G,\bu v_2),\tpp{f(q,l)}(G,\bu v_1))$.
    W.l.o.g.\ we assume it is the former form.

    We show that $R_1 \cap R_2 \subseteq \elements(\bu)$.
    To this end, consider any vertex $u' \in R_1 \cap R_2$ and corresponding hidden vertex $h'$.
    If $\sigma(u') > \sigma(u)$, then this would mean that $\pi(h') > \pi(h)$,
    a contradiction to the fact that we chose $h$ to be the largest vertex 
    in $\bigl(N^-(v_1) \cap N^+(v_2) \bigr) \cup \bigl(N^+(v_1) \cap N^-(v_2)\bigr)$.
    Therefore $\sigma(u') \le \sigma(u)$.
    We constructed $L$ such that $u \in \wreach_{4^{q}}(G,\sigma,v_1)$
    and we chose $u' \in \wreach_{4^{q}}(G,\sigma,v_1)$.
    Thus by \myitem{2r} of \autoref{prop:colfacts}, $u' \in \wreach_{2 \cdot 4^{q}}(G,\sigma,u) = \elements(\bu)$.
    This implies that $R_1 \cap R_2 \subseteq \elements(\bu)$.

    Thus, \myitem{rpath} of \autoref{prop:colfacts} states that $\bu$ $4^q$-separates $v_1$ and $v_2$.
    Hence, as discussed earlier, the fact whether $G \models \phi(v_1,v_2) \lor \phi(v_2,v_1)$ only depends on $\tpp{f(q,l)}(G,\bu v_1)$ and $\tpp{f(q,l)}(G,\bu v_2)$.
    We constructed $L$ such that
    $h = (u,\bu,\tpp{f(q,l)}(G,\bu v_1),\tpp{f(q,l)}(G,\bu v_2))$ 
    is labeled with ``1'' if and only if $G \models \phi(v_1,v_2) \lor \phi(v_2,v_1)$, i.e., $v_1v_2 \in E(I_\phi(G))$.
    This implies that $(L,\pi)$ is in fact a directed lacon-decomposition of $I_\phi(G)$.

\paragraph{Bounding the Coloring Numbers.}
    For a hidden vertex $(u,\bu,\typi_1,\typi_2)$ in $L$, we say $u$ is its \emph{corresponding vertex in $G$}.
    If $t$ is a target vertex, we say the \emph{corresponding vertex in $G$} is $t$ itself.
    The corresponding vertex of a vertex $x \in V(L)$ is denoted by $u(x)$.

    We start by bounding the number of hidden vertices that have the same corresponding vertex in $G$.
    To do so, we need to determine the number of possible types $\typi_1,\typi_2$ a hidden vertex can have.
    The size of $\bu$ is bounded by $\wcol_{2\cdot 4^q}(G,\sigma)$.
    Thus, the number of $f(q,l)$-types containing formulas with at most $|\bu|+1$ free variables is bounded by some function
    of $q$ and $l$.
    This means, there exists a function $g'$ (independent of $G$ and $\sigma$) such~that
    \begin{equation}\label{eq:samecorrvertex}
    \text{ for all $v \in V(G)$ holds } |\{ x \in V(L) \mid u(x) = v \}| \le g'(q,l).
    \end{equation}

    Let $h$ be a hidden vertex and $h' \in \reach_{r}(L,\pi,h)$ for some $r$.
    Since all target vertices are larger than $h$ with respect to $\pi$, we know that $h'$ is a hidden vertex.
    Remember that $L$ is bipartite.
    By definition of the generalized coloring numbers, there is a path $h_0 t_0 h_1 t_1 \dots t_{k-1} h_k$ such that
    $h_0 = h'$, $h_k=h$, $k \le r/2$,
    $\pi(h_0) < \pi(h_k)$, and $\pi(h_k) = \min_{i=2}^k \pi(h_i)$.
    Here, the $h_i$ are hidden vertices and the $t_i$ are target vertices.
    The order $\pi$ was constructed such that $\sigma(u(h_i)) > \sigma(u(h_j))$ implies $\pi(h_i) > \pi(h_j)$ for all hidden vertices $h_i,h_j$.
    Therefore, $\sigma(u(h_0)) \le \sigma(u(h_k))$, and $\sigma(u(h_k)) = \min_{i=2}^k \sigma(u(h_i))$.
    We constructed $L$ such that $u(h_i),u(h_{i+1}) \in \wreach_{4^q}(G,\sigma,t_i)$.
    Thus by \myitem{2r} of \autoref{prop:colfacts}, 
    either $u(h_i) \in \wreach_{2 \cdot 4^q}(G,\sigma,u(h_{i+1}))$ or $u(h_{i+1}) \in \wreach_{2 \cdot 4^q}(G,\sigma,u(h_{i}))$.
    By \myitem{bugfix-kr} of \autoref{prop:colfacts}, 
    there exists $w \in \reach_{4^q r}(G,\sigma,u(h))$ with $u(h') \in \wreach_{2 \cdot 4^q}(G,\sigma,w)$.
    In other words, 
    $|\reach_{r}(L,\pi,h)|$ is bounded from above by the number of tuples $(w,v,h')$ with
    $w \in \reach_{4^q r}(G,\sigma,u(h))$, $v \in \wreach_{2 \cdot 4^q}(G,\sigma,w)$ and $u(h') = v$.
    Using (\ref{eq:samecorrvertex}), we bound the number of such tuples by
    \begin{equation}\label{eq:boundcolhidden}
        |\reach_{r}(L,\pi,h)| \le \col_{4^q r}(G,\sigma) \cdot l \cdot g'(q,l).
    \end{equation}

    Let us now consider a target vertex $t$.
    Observe that
    \begin{equation}\label{eq:boundcoltarget1}
        |\reach_{r}(L,\pi,t)| \le 1 + \sum_{h \in N(t)\setminus \{t\}} |\reach_{r}(L,\pi,h)|.
    \end{equation}
    Since for all $h \in N(t)$ holds $u(h) \in \wreach_{4^q}(G,\sigma,t)$,
    we can again use (\ref{eq:samecorrvertex}) to bound
    \begin{equation}\label{eq:boundcoltarget2}
        |N(t)| \le g'(q,l) \cdot l.
    \end{equation}
    By \myitem{colwcol} of \autoref{prop:colfacts}, $l \le \col_{2\cdot 4^q}(G,\sigma)^{2 \cdot 4^q}+2$.
    Combining equations (\ref{eq:boundcolhidden}), (\ref{eq:boundcoltarget1}) and (\ref{eq:boundcoltarget2}) yields a function $g$ with
    \[
    \col_r(L,\pi) \le g(q + \col_{2 \cdot 4^{q}}(G,\sigma)) \cdot \col_{4^{q} r}(G,\sigma) \text{ for all $r$}.
    \]

    The statement of this lemma also requires a bound on the weak coloring numbers $\wcol_r(L,\pi)$.
    This bound can be proved in almost the same way as for $\col_r(L,\pi)$ and therefore we only describe the main difference.
    Let $h$ be a hidden vertex and $h' \in \wreach_{r}(L,\pi,h)$.
    By definition, there is a path $h_0 t_0 h_1 t_1 \dots t_{k-1} h_k$ such that
    $h_0 = h'$, $h_k=h$, $k \le r/2$,
    $\pi(h_0) = \min_{i=1}^k \pi(h_i)$.
    Then by \autoref{prop:colfacts}, Item~\myitem{weak-kr},
    $u(h') \in \wreach_{4^q r}(G,\sigma,u(h))$,
    i.e., 
    $|\wreach_{r}(L,\pi,h)|$ is bounded by the number of tuples $(v,h')$ with
    $v \in \wreach_{4^q r}(G,\sigma,u(h))$ and $u(h') = v$.
    The rest proceeds as for the strong coloring numbers.

\paragraph{No Apex Vertex.}
    It remains to prove this result for the case that the input graph has no apex vertex.
    We reduce this case to the previously covered case with an apex.
    Let $G$ be a graph without an apex and $\sigma$ be an ordering of $V(G)$.
    We construct a graph $G'$ from $G$ by adding an additional apex vertex and let $\sigma'$ be the ordering of $V(G')$
    that preserves the order of $\sigma$ but whose minimal element is the new apex vertex.
    Then
    $\col_r(G',\sigma') \le \col_{r}(G,\sigma)+1$ and 
    $\wcol_r(G',\sigma') \le \wcol_{r}(G,\sigma)+1$.
    Assume we have a directed lacon-decomposition $(L',\pi')$ of $I_\phi(G')$.
    We obtain a directed lacon-decomposition $(L,\pi)$ of $I_\phi(G)$ by removing the apex vertex from $(L',\pi')$.
    Then
    $\col_r(L,\pi) \le \col_{r}(L',\pi')$ and
    $\wcol_r(L,\pi) \le \wcol_{r}(L',\pi')$.
    These observations reduce the no-apex case to the apex case.
\end{proof}

\section{Directed and Undirected Lacon-Decompositions Have\texorpdfstring{\\}{ }Same Expressive Power}\label{sec:dirtoundir}

In this section, we prove the following lemma, stating that directed and normal lacon-decompositions are equally powerful.

\begin{lem}\label{lem:dirtoundir}
    Assume a graph $G$ has a directed lacon-decomposition $(L,\pi)$.
    Then it also has an (undirected) lacon-decomposition $(L',\pi')$ with
    \begin{itemize}
        \item $\col_r(L',\pi') \le  4^{\col_2(L,\pi)} \cdot  \col_r(L,\pi)$ for all~$r$,
        \item $\wcol_r(L',\pi') \le 4^{\col_2(L,\pi)} \cdot \wcol_r(L,\pi)$ for all~$r$.
    \end{itemize}
\end{lem}

\begin{proof}
Assume we have a directed lacon-decomposition $(L,\pi)$ of a graph $G$ with target vertices $T$ and hidden vertices $H$.
We construct an undirected lacon-decomposition $(L',\pi')$ of $G$ with target and hidden vertices $T,H'$ as follows:
We start with $H' = \emptyset$.
We process the vertices in $H$ one by one in ascending order by $\pi$ and add in each processing step various new vertices to $H'$.
The ordering $\pi'$ is thereby defined implicitly as follows (in ascending order):
First come the vertices of $H'$ in the order of insertion and then come the vertices of $T$ in the same order as in $\pi$.

We will make sure that after a vertex $h \in H$ has been processed the following invariant holds for all pairs of vertices $v_1 \neq v_2 \in T$.
If $v_1,v_2$ have a dominant vertex $d$ in $L$ such that $\pi(d) \le \pi(h)$,
then we guarantee that $v_1,v_2$ have a dominant vertex $d'$ in $L'$
and that $d'$ is labeled with ``1'' if and only if $d$ is.
After every vertex in $H$ has been processed,
this guarantees that $(L',\pi')$ is an undirected lacon-decomposition of $G$.

While processing a vertex $h \in H$, certain new vertices are inserted into $H'$.
The newly inserted vertices are in a sense either a copy of $h$ or copies
of vertices that have already been inserted into $H'$.
These copies may later again be copied, and so forth.
This leads to exponential growth.
To keep track of these copies and to show that while being exponential, this grow is nevertheless bounded,
every newly inserted vertex $h'$ will have a \emph{memory}, denoted by $m(h')$.
This memory consists of a sequence of vertices from $H$.
If $h'$ is derived from $h$, then we set $m(h') = h$.
However, if $h'$ is derived both from $h$ and another vertex $h''$ that has previously been inserted into $H'$, then we set $m(h') = m(h'')+h$, i.e.,
$m(h')$ is obtained by appending $h$ to the memory of $h''$.
Let us now get into the details of the construction
and prove our invariant.

\paragraph{Construction of Lacon-Decomposition.}
    We describe the processing steps of a vertex $h \in H$.
    If $N^-(h) \cup N^+(h) = \emptyset$ we do nothing. 
    Otherwise, we add a vertex $h'$ to $H'$ that has the same label (``0'' or ``1'')  as $h$. We further add edges to $L'$ such that $N(h') = N^-(h) \cup N^+(h)$
    (here $N(h')$ refers to the neighborhood in $L'$, while $N^-(h)$, $N^+(h)$ refer to the in- and out-neighborhoods in $L$).
    Since $h'$ is derived only from $h$, we set $m(h') = h$.

    We observe a problem.
    Let $v_1 \neq v_2 \in N^-(h) \setminus N^+(h)$.
    Currently, $h'$ is the dominant vertex for $v_1$ and $v_2$ in $L'$, even though $h$ is not dominant for these vertices in $L$.
    Thus $h'$ connects too many target vertices and we have to ``undo'' the effect of $h'$ on the edges within the set $N^-(h)\setminus N^+(h)$.
    The same holds for the set $N^+(h)\setminus N^-(h)$.
    We do so as follows.
    We iterate over all vertices $l \in \reach_2(L',\pi',h')$ in order of insertion into $H'$.
    If $N(l) \cap N^+(h)\setminus N^-(h) \neq \emptyset$ we add a vertex $l'$ with $N(l') = N(l) \cap N^+(h)\setminus N^-(h)$.
    Similarly,
    if $N(l) \cap N^-(h)\setminus N^+(h) \neq \emptyset$ we add a vertex $l''$ with $N(l'') = N(l) \cap N^-(h)\setminus N^+(h)$.
    These two vertices get the same label as $l$ and since they were derived from both $l$ and $h$, we set
    $m(l') = m(l'') = m(l)+h$.
    These new vertices $l'$ and $l''$ undo the undesired effects of inserting $h'$, as we will prove soon.
    This completes the processing round of $h$.

    By induction,
    we know that the invariant was satisfied in the last round for all $v_1,v_2$.
    We show that it also holds after $h$ is processed.
    Let $v_1 \neq v_2$ be two vertices with dominant vertex $d$ in $L$ such that $\pi(d) \le \pi(h)$.
    We distinguish five cases.
    \begin{itemize}
        \item $v_1 \not\in N^-(h) \cup N^+(h)$ or $v_2 \not\in N^-(h) \cup N^+(h)$. 
            In this case, neither $h$ in $(L,\pi)$ nor any of the newly inserted vertices in $(L',\pi')$ is dominant for $v_1,v_2$.
            Since the invariant for $v_1,v_2$ held in the previous round, it also holds in this round.
            
        \item $v_1 \in N^-(h)$ and $v_2 \in N^+(h)$.
            In this case, $h$ and $h'$ are dominant for $v_1,v_2$ in $(L,\pi)$ and $(L',\pi')$, respectively.
            The vertex $h'$ is labeled with ``1'' if and only if $h$ is.
            Thus, the invariant is fulfilled.

        \item $v_1 \in N^+(h)$ and $v_2 \in N^-(h)$. As above.

        \item $v_1, v_2 \in N^+(h) \setminus N^-(h)$.
            Then $\pi(d) < \pi(h)$.
            Thus, in the previous round, there was a vertex that was dominant for $v_1,v_2$ in $(L',\pi')$ that has the same label as $d$.
            Since $v_1,v_2 \in N(h')$, this dominant vertex is contained in $\reach_2(L',\pi',h')$.
            In the construction, we iterate over all vertices $l \in \reach_2(L',\pi',h')$ in order of insertion into $H'$ (i.e., ascending order by $\pi'$)
            and insert vertices $l'$ that satisfy $v_1,v_2 \in N(l')$ iff $v_1,v_2 \in N(l)$.
            The last vertex $l$ with $v_1,v_2 \in N(l)$ that we encounter during this procedure
            was the dominant vertex of the previous round.
            The corresponding vertex $l'$ has the same label as $l$ and is the dominant vertex now.

        \item $v_1, v_2 \in N^-(h) \setminus N^+(h)$.
            As above, but with $l''$ instead of $l'$.
    \end{itemize}

\paragraph{Bounding the Coloring Numbers.}
For a vertex $h' \in H'$ with $m(h') = h_1 \dots h_k$, we say that $h_k$ is \emph{the corresponding vertex of $h'$ in $L$}.
If $t$ is a target vertex, we say the \emph{corresponding vertex in $L$} is $t$ itself.
The corresponding vertex of a vertex $x \in V(L')$ is denoted by $u(x)$.

We fix a vertex $h \in H$ and ask: How many $h' \in H'$ can there be with $u(h')~=~h$?
We pick a vertex $h' \in H'$ with $u(h') = h$.
Then it has a memory $m(h') = h_1\dots h_k$ with $h_k = h$.
The following observations follow from the construction of $(L',\pi')$.
\begin{itemize}
    \item $N(h') \neq \emptyset$
    \item $N(h') \subseteq N^-(h_i) \cup N^+(h_i)$ for all $1 \le i \le k$
    \item $\pi(h_1) < \dots < \pi(h_k)$
\end{itemize}
These three observations together imply that $h_1,\dots,h_k \in \reach_2(L,\pi,h)$.
Thus $m(h')$ consists of a subset of $\reach_2(L,\pi,h)$, written in ascending order by $\pi$.
We can therefore bound the number of possible memories $m(h')$ of $h'$ by $2^{\col_2(L,\pi)}$.

Furthermore, for each vertex $l \in H'$ with memory $l_1\dots l_{k-1}$ and every vertex $l_k \in H$ there are at most two vertices
(denoted by $l'$ and $l''$ in the above construction) with the memory $l_1\dots l_k$.
This means for a fixed memory of length $k$, there are at most $2^k$ vertices that have this particular memory.
Combining the previous observations, we can conclude that 
for a fixed vertex $h \in H$ there are at most $2^{\col_2(L,\pi)} \cdot 2^{\col_2(L,\pi)} = 4^{\col_2(L,\pi)}$ vertices $h' \in H'$ with $u(h') = h$.

The fact that $N(h') \subseteq N^-\bigl(u(h')\bigr) \cup N^+\bigl(u(h')\bigr)$for all $h' \in H'$ (see second item in enumeration above) implies the following for all $x,y \in V(L')$:
If $x,y$ are adjacent in $L'$, then $u(x),u(y)$ are adjacent in $L$.
Also, the order $\pi'$ was constructed such that $\pi(u(x)) > \pi(u(y))$ implies $\pi'(x) > \pi'(y)$ for all $x,y \in V(L')$.
The last two points together mean
\begin{itemize}
    \item if $y \in \reach_{r}(L',\pi',x)$,  then $u(y) \in \reach_{r}(L,\pi,u(x))$ for all $r$,
    \item if $y \in \wreach_{r}(L',\pi',x)$, then $u(y) \in \wreach_{r}(L,\pi,u(x))$ for all $r$.
\end{itemize}
Combining this with the observation that
for a fixed $u \in V(L)$ there are at most $4^{\col_2(L,\pi)}$ vertices $y \in V(L')$ with $u(y) = u$
gives us the bounds on  
$\col_r(L',\pi')$
and $\wcol_r(L',\pi')$
required by this lemma.
\end{proof}

\section{Transductions of Bounded Expansion Have Lacon-Decompositions}\label{sec:onetothree}

Now we combine the previous results into a more concise statement.
We show that for every transduction of a graph $G$, we can find
a lacon-decomposition whose coloring numbers are not too far off from the coloring
numbers of $G$.

\begin{thm}\label{thm:maindirection}
Let $\tau$ be a transduction.
There exist a constant $c$ and a function $f$ such that
for every graph $G$, ordering $\sigma$ on the vertices of $G$, and $D \in \tau(G)$
there exists a lacon-decomposition $(L,\pi)$ of $D$ with
\begin{itemize}
    \item $\col_r(L,\pi)  \le f(\col_{c}(G,\sigma)) \cdot \col_{cr}(G,\sigma)$ for all $r$,
    \item $\wcol_r(L,\pi) \le f(\col_{c}(G,\sigma)) \cdot \wcol_{cr}(G,\sigma)$ for all $r$.
\end{itemize}
\end{thm}

All the key ideas needed to prove this result have already been established in \autoref{lem:interpretationtolacon} and \autoref{lem:dirtoundir}.
In this section, we merely combine them.
First, we need to following technical but unexciting lemma.
\begin{lem}\label{lem:basictransduction}
    Let $\tau$ be a transduction.
    There exists a constant $c$ and a basic transduction $\tau_0$ such that for every
    graph $G$, ordering $\pi$ on the vertices of $G$, and $D \in \tau(G)$
    there exists a labeled graph $G'$ with $\tau_0(G') = D$ and ordering $\pi'$ on the vertices of $G'$ such that
\begin{itemize}
    \item $\col_r(G',\pi') \le   c \cdot  \col_r(G,\pi)$ for all $r$,
    \item $\wcol_r(G',\pi') \le  c \cdot \wcol_r(G,\pi)$ for all $r$.
\end{itemize}
\end{lem}
\begin{proof}
    We break $\tau$ down into $\tau = \tau_0 \circ \gamma \circ \epsilon$.
    Let $G' \in (\gamma \circ \epsilon)(G)$ such that $\tau_0(G') = D$.
    Note that $G'$ is the result of applying the copy operation $\epsilon$ and a coloring from $\gamma$ to $G$.
    We extend $\pi$ into an ordering $\pi'$ on $G'$ by placing the copied vertices next to the corresponding original vertices.
    If $\epsilon$ copies the graph $c$ times, then
    $\col_r(G',\pi')  \le c \cdot   \col_r(G,\pi)$ and
    $\wcol_r(G',\pi') \le c \cdot  \wcol_r(G,\pi)$.
\end{proof}

Now we prove \autoref{thm:maindirection}
by chaining \autoref{lem:basictransduction}, \ref{lem:interpretationtolacon} and \ref{lem:dirtoundir} in this order.

\begin{proof}[Proof of \autoref{thm:maindirection}]
    Assume we have a graph $G$, an ordering $\sigma$ on the vertices of $G$, and $D \in \tau(G)$.
    \autoref{lem:basictransduction} gives us a basic transduction $\tau_0$, a graph $G'$ with $\tau_0(G') = D$ and an ordering $\sigma'$ such that
    \begin{itemize}
        \item $\col_r(G',\sigma') \le   c' \cdot  \col_r(G,\sigma)$ for all $r$,
        \item $\wcol_r(G',\sigma') \le  c' \cdot \wcol_r(G,\sigma)$ for all $r$.
    \end{itemize}
    for some constant $c'$ depending only on $\tau$.
    Assume $\tau_0$ to be of the form $(\chi,\nu,\phi)$.
    By \autoref{lem:interpretationtolacon},
    there exists a function $g$ and a directed lacon-decomposition $(L',\pi')$ of $I_\phi(G')$ with
    \begin{itemize}
        \item $\col_r(L',\pi') \le g(|\phi| + \col_{2 \cdot 4^{|\phi|}}(G',\sigma')) \cdot \col_{4^{|\phi|} r}(G',\sigma')$ for all $r$,
        \item $\wcol_r(L',\pi') \le g(|\phi| + \col_{2 \cdot 4^{|\phi|}}(G',\sigma')) \cdot \wcol_{4^{|\phi|} r}(G',\sigma')$ for all $r$.
    \end{itemize}
    We know $G' \models \chi$, since otherwise $\tau_0(G')$ would be undefined.
    Since $\nu$ describes the vertex set of the transduction,
    we update $L'$ by removing all target vertices $t$ with $L' \not\models \nu(t)$.
    $(L',\pi')$ is now a directed lacon-decomposition of $\tau_0(G') = D$.
    Since we only deleted vertices, the coloring numbers of $(L',\pi')$ did not increase, thus the above two bounds remain true.
    At last, we use \autoref{lem:dirtoundir} to construct an undirected lacon-decomposition $(L,\pi)$ of $D$ with
    \begin{itemize}
        \item $\col_r(L,\pi) \le   4^{\col_2(L',\pi')}  \cdot \col_r(L',\pi')$ for all $r$,
        \item $\wcol_r(L,\pi) \le  4^{\col_2(L',\pi')} \cdot \wcol_r(L',\pi')$ for all $r$.
    \end{itemize}
    Combining the previous 6 bounds then proves the result.
\end{proof}

\section{Converting Lacon- to Shrub-Decompositions}\label{sec:lacontoshrub}

We now convert a lacon-decomposition into an equivalent shrub-decomposition while maintaining a bound on the generalized coloring numbers.

\begin{lem}\label{lem:lacontoshrub}
    Let $G$ be a graph with a lacon-decomposition $(L,\pi)$.
    Then there exists a shrub-decomposition $(F,S)$ of $G$ with one color, diameter at most $12\col_1(L,\pi)+2$ and
    \begin{itemize}
        \item $\col_r(F) \le  7\col_1(L,\pi) + 1 + \col_r(L,\pi)$ for all $r$,
        \item $\wcol_r(F) \le 7\col_1(L,\pi) + 1 + \wcol_r(L,\pi)$ for all $r$.
    \end{itemize}
\end{lem}
\begin{proof}
    Let $(L,\pi)$ be a lacon-decomposition of $G$.
    For every target vertex $t$ and $1 \le i < |N(t)|$
    we define $\nu_i(t)$ to be the $i$th hidden neighbor in $L$, counted in descending order by $\pi$.
    We construct a graph $F$ from $L$ (see \Cref{fig:lacontoshrub}) by subdividing for every \(t \in V(G)\) and \(1 \le i \le |N(t)|\) 
    the edge between \(t\) and \(\nu_i(t)\) into a path of length
    \[
    \begin{cases}
        4\col_1(L,\pi)+2i     & \textnormal{ if \(\nu_i(t)\) is labeled with ``0''}, \\
        4\col_1(L,\pi)+2i + 1 & \textnormal{ if \(\nu_i(t)\) is labeled with ``1''}.
    \end{cases}
    \]
    We furthermore iteratively remove pendant vertices in $V(F)\setminus V(G)$ from $F$ until the pendant vertices are exactly $V(G)$.

    Consider a path between two distinct vertices $t_1, t_2 \in V(G)$ in \(F\).
    If it passes through two hidden vertices, then it has length at least \(4\cdot 4\col_1(L,\pi)\), 
    while a path passing through only one hidden vertex has length at most \(2\cdot (4\col_1(L,\pi) + 2\col_1(L,\pi)+1) = 12\col_1(L,\pi)+2\), which is shorter.
    Thus, the shortest path between any two vertices $t_1, t_2 \in V(G)$ in \(F\) passes through exactly one hidden vertex, namely their largest common neighbor in $L$.
    Each half of this path has even length iff the largest common neighbor is labeled with~``0''.
    Thus, two vertices are adjacent in $G$ iff their distance in $F$ modulo four is two.

\begin{figure}
\begin{center}
\def\ty{2}
\def\tx{2}
\scalebox{0.9}{
\begin{tikzpicture}[align=center,scale=1] 
    \def\borderthickness{1}
    \tikzstyle{target}=[draw=black]
    \tikzstyle{circ}=[circle,draw=black, inner sep=0.05cm]
    \tikzstyle{dot}=[circle,draw=black,fill=black,minimum size=0.2cm, inner sep=0cm]
    \tikzstyle{edge}=[draw=black, line width=\borderthickness]

    \node[target,fill=black,text=white] at (-1.5*\tx,0)                     (a) {1}; 
    \node[target,fill=white] at (-0.5*\tx,0)                     (b) {0}; 
    \node[target,fill=black,text=white] at ( 0.5*\tx,0)                     (c) {1}; 
    \node[target,fill=white] at ( 1.5*\tx,0)                     (d) {0}; 

    \node[circ,fill=gray!40] at (-0.5*\tx,-\ty)                     (t1) {$t_1$}; 
    \node[circ,fill=gray!40] at (+0.5*\tx,+\ty)                     (t2) {$t_2$}; 

    \draw[edge] (t1) to (a);
    \draw[edge] (t1) to (b);
    \draw[edge] (t1) to (c);
    \draw[edge] (t2) to (b);
    \draw[edge] (t2) to (c);
    \draw[edge] (t2) to (d);
\end{tikzpicture}
\hspace{2.1cm}
\begin{tikzpicture}[align=center,scale=1] 
    \def\borderthickness{1}
    \tikzstyle{target}=[draw=black]
    \tikzstyle{circ}=[circle,draw=black, inner sep=0.05cm]
    \tikzstyle{dot}=[circle,draw=black,fill=black,minimum size=0.1cm, inner sep=0cm]
    \tikzstyle{edge}=[draw=black, line width=\borderthickness]

    \node[target,fill=black,text=white] at (-1.5*\tx,0)                     (a) {1}; 
    \node[target,fill=white] at (-0.5*\tx,0)                     (b) {0}; 
    \node[target,fill=black,text=white] at ( 0.5*\tx,0)                     (c) {1}; 
    \node[target,fill=white] at ( 1.5*\tx,0)                     (d) {0}; 

    \node[circ,fill=gray!40] at (-0.5*\tx,-\ty)                     (t1) {$t_1$}; 
    \node[circ,fill=gray!40] at (+0.5*\tx,+\ty)                     (t2) {$t_2$}; 

    \draw[edge] (t1) edge node{\small $46{+}1$\hspace{1.1cm}} (a);
    \draw[edge] (t1) edge node{\small \hspace{0.9cm}$44{+}0$} (b);
    \draw[edge] (t1) edge node{\small \hspace{1.1cm}$42{+}1$} (c);
    \draw[edge] (t2) edge node{\small $46{+}0$\hspace{1.1cm}} (b);
    \draw[edge] (t2) edge node{\small \hspace{0.9cm}$44{+}1$} (c);
    \draw[edge] (t2) edge node{\small \hspace{1.1cm}$42{+}0$} (d);
\end{tikzpicture}
}
\caption{%
    Left: Part of a lacon-decomposition.
    Right: Corresponding part of a shrub-decomposition, including
    distances between round and square vertices.
    We assume \(4\col_1(L,\pi)= 40\).
}
\label{fig:lacontoshrub}
\end{center}
\end{figure}
    
    Since our construction does not rely on any colors, we give every target vertex the same dummy color $1$.
    The diameter of $F$ is at most $12\col_1(L,\pi)+2$.
    We define the signature of the shrub decomposition as
    \[S = \bigl\{
    (1,1,4d+2) \mid 1 \le d \le 3\col_1(L,\pi) \bigr\}.
    \]
    The bound on the coloring numbers of $F$ can be proved as follows.
    Let $H$ be the hidden vertices of $L$.
    Every connected component $X$ of $F[V(F) \setminus H]$
    consists of at most \(\col_1(L,\pi)\) many paths of length at most \(6\col_1(L,\pi)\), intersecting in some shared vertex \(t \in V(G)\).
    Thus, \(X\) has size at most \(7\col_1(L,\pi)+1\).

    We extend the ordering $\pi$ of $L$ into an ordering $\pi'$ of $F$
    by making the vertices in $V(F) \setminus V(L)$ larger than all vertices in $V(L)$.
    Now for every $h \in H$ we have
    $\reach_r(F,\pi',h) \subseteq \reach_r(L,\pi,h)$.
    For all other vertices $v \in V(F) \setminus H$ holds
    $\reach_r(F,\pi',v) \subseteq \reach_r(L,\pi,t) \cup X$,
    where $X$ is the connected component of $v$ in $F[V(F) \setminus H]$ (of size at most \(7\col_1(L,\pi)+1\))
    and $t \in V(G)$ is the unique target vertex contained in $X$.
    An equivalent bound can be obtained for the weak reachability.
\end{proof}

\section{Converting Lacon- to Parity-Decompositions}\label{sec:lacontoparity}

The following proof is substantially similar to the one of \autoref{lem:dirtoundir}.

\begin{lem}\label{lem:lacontoparity}
    Assume a graph $G$ has a lacon-decomposition $(L,\pi)$.
    Then it also has a parity-decomposition $P$ with target-degree at most $2^{\col_2(L,\pi)} \cdot  \col_1(L,\pi)$ and
    \begin{itemize}
        \item $\col_r(P) \le  2^{\col_2(L,\pi)} \cdot  \col_r(L,\pi)$ for all~$r$,
        \item $\wcol_r(P) \le 2^{\col_2(L,\pi)} \cdot \wcol_r(L,\pi)$ for all~$r$.
    \end{itemize}
\end{lem}

\begin{proof}
Assume we have a lacon-decomposition $(L,\pi)$ of a graph $G$ with target vertices $T$ and hidden vertices $H$.
We construct a parity-decomposition $P$ of $G$ with target and hidden vertices $T,H'$ as follows:
We start with $H' = \emptyset$.
We process the vertices in $H$ one by one in ascending order by $\pi$ and add in each processing step various new vertices to $H'$.
This defines an ordering $\pi'$ of $V(P)$ as follows (in ascending order):
First come the vertices of $H'$ in the order of insertion and then come the vertices of $T$ in the same order as in $\pi$.
Afterwards, we will use $\pi$ to bound the coloring numbers of~$P$.

We will make sure that after a vertex $h \in H$ has been processed the following invariant holds for all pairs of vertices $v_1 \neq v_2 \in T$.
If $v_1,v_2$ have a dominant vertex $d$ in $L$ such that $\pi(d) \le \pi(h)$
then we guarantee that 
$|N(v_1) \cap N(v_2)|$ is odd in $P$
if and only if $d$ labeled with ``1''.
After every vertex in $H$ has been processed,
this guarantees that $P$ is a parity-decomposition of $G$.

While processing a vertex $h \in H$, certain new vertices are inserted into $H'$.
The newly inserted vertices are in a sense either a copy of $h$ or copies
of vertices that have already been inserted into $H'$.
These copies may later again be copied, and so forth.
This leads to exponential growth.
To keep track of these copies and to show that while being exponential, this grow is nevertheless bounded,
every newly inserted vertex $h'$ will have a \emph{memory}, denoted by $m(h')$.
This memory consists of a sequence of vertices from $H$.
If $h'$ is derived from $h$, then we set $m(h') = h$.
However, if $h'$ is derived both from $h$ and another vertex $h''$ that has previously been inserted into $H'$, then we set $m(h') = m(h'')+h$, i.e.,
$m(h')$ is obtained by appending $h$ to the memory of $h''$.
Let us now get into the details of the construction
and prove our invariant.

\paragraph{Construction of Parity-Decomposition.}
We describe the processing steps of a vertex $h \in H$.
If $N^L(h) = \emptyset$ we do nothing. 
Otherwise iterate over all $l \in H'$ with $N^{P}(l) \cap N^{L}(h) \neq \emptyset$ in order of insertion to $H'$,
and add a vertex $l'$ to $P$ with $N^{P}(l') = N^{P}(l) \cap N^L(h)$ and $m(l') = m(l)+h$.
Also, if $h$ is labeled with ``1'', we add an additional vertex $h'$ to $P$ with $N^{P}(h')=N^L(h)$ and $m(h')=h$. 
By induction,
we know that the invariant was satisfied in the last round for all $v_1,v_2$.
We show that it also holds after $h$ is processed.
Let $v_1 \neq v_2$ be two vertices with dominant vertex $d$ in $L$ such that $\pi(d) \le \pi(h)$.
We distinguish two cases.
\begin{itemize}
    \item $v_1 \not\in N^L(h)$ or $v_2 \not\in N^L(h)$.
        In this case, $h$ is not dominant for $v_1,v_2$ in $(L,\pi)$. Also none of the newly inserted vertices are contained in $|N^{P}(v_1) \cap N^{P}(v_2)|$.
        Since the invariant for $v_1,v_2$ held in the previous round, it also holds in this round.
        
    \item $v_1,v_2 \in N^L(h)$.
        Now, $h$ is dominant for $v_1,v_2$ in $(L,\pi)$.
        If $h$ is labeled with ``0'', then $|N^{P}(v_1) \cap N^{P}(v_2)|$ is even.
        This is because for every element $l \in N^{P}(v_1) \cap N^{P}(v_2)$ from the previous round,
        a ``copy'' $l'$ with $v_1,v_2 \in N^{P}(l')$ has been inserted.
        On the other hand,, if $h$ is labeled with ``1'', the additional vertex $h'$ with $N^{P}(h')=N^L(h)$ makes 
        $|N^{P}(v_1) \cap N^{P}(v_2)|$ odd.
\end{itemize}

\paragraph{Bounding the Coloring Numbers.}
For a vertex $h' \in H'$ with $m(h') = h_1 \dots h_k$, we say that $h_k$ is \emph{the corresponding vertex of $h'$ in $L$}.
If $t$ is a target vertex, we say the \emph{corresponding vertex in $L$} is $t$ itself.
The corresponding vertex of a vertex $x \in V(P)$ is denoted by $u(x)$.

We fix a vertex $h \in H$ and ask: How many $h' \in H'$ can there be with $u(h') = h$?
We pick a vertex $h' \in H'$ with $u(h') = h$.
Then it has a memory $m(h') = h_1\dots h_k$ with $h_k = h$.
The following observations follow from the construction of $P$.
\begin{itemize}
    \item $N^P(h') \neq \emptyset$
    \item $N^P(h') \subseteq N^L(h_i)$ for all $1 \le i \le k$
    \item $\pi(h_1) < \dots < \pi(h_k)$
\end{itemize}
These three observations together imply that $h_1,\dots,h_k \in \reach_2(L,\pi,h)$.
Thus $m(h')$ consists of a non-empty subset of $\reach_2(L,\pi,h)$, written in ascending order by $\pi$.
We can therefore bound the number of possible memories $m(h')$ of $h'$ by $2^{\col_2(L,\pi)}-1$.
Also, every vertex $h' \in H'$ is uniquely identified by its memory $m(h')$ and every vertex $t \in T$ has $u(t)=t$.
We can therefore conclude
for for every $h \in H$ that there are at most $2^{\col_2(L,\pi)}$ vertices $h' \in V(P)$ with $u(h') = h$.

The fact that $N^P(h') \subseteq N^L\bigl(u(h')\bigr)$ for all $h' \in H'$ (see second item in enumeration above) implies the following for all $x,y \in V(P)$:
If $x,y$ are adjacent in $P$, then $u(x),u(y)$ are adjacent in $L$.
Also, the order $\pi'$ was constructed such that $\pi(u(x)) > \pi(u(y))$ implies $\pi'(x) > \pi'(y)$ for all $x,y \in V(P)$.
The last two points together mean
\begin{itemize}
    \item if $y \in \reach_{r}(P,\pi',x)$,  then $u(y) \in \reach_{r}(L,\pi,u(x))$ for all $r$,
    \item if $y \in \wreach_{r}(P,\pi',x)$, then $u(y) \in \wreach_{r}(L,\pi,u(x))$ for all $r$.
\end{itemize}
Combining this with the observation that
for a fixed $u \in V(L)$ there are at most $2^{\col_2(L,\pi)}$ vertices $y \in V(P)$ with $u(y) = u$
gives us the bounds on  
$\col_r(P,\pi')$
and $\wcol_r(P,\pi')$
required by this lemma.
Since every target vertex in $L$ has degree at most $\col_1(L,\pi)$,
$P$ has target-degree at most $2^{\col_2(L,\pi)} \cdot \col_1(L,\pi)$.
\end{proof}

\section{Proof of \autoref{thm:BEcharacterization}, \ref{thm:TDcharacterization} and \ref{thm:TWcharacterization}}\label{sec:threetheorems}

We finally prove the main result of this paper by combining all results of the previous sections.

\begin{proof}[Proof of \autoref{thm:BEcharacterization}]
    We prove
    \myitem{bebase} $\Longrightarrow$ \myitem{belacon} $\Longrightarrow$ \myitem{beshrub} $\Longrightarrow$ \myitem{bebase}
    and then \myitem{belacon} $\Longrightarrow$ \myitem{beparity} $\Longrightarrow$ \myitem{bebase}.
    \begin{itemize}
        \item \myitem{bebase} $\Longrightarrow$ \myitem{belacon}.
            Pick any graph $D \in \G$. By \myitem{bebase} there exists a graph $G \in \G'$ with $D \in \tau(G)$.
            Since $\G'$ has bounded expansion, \autoref{def:be} states that there exists
            an ordering $\sigma$ with $\col_r(G,\sigma) \le g(r)$ for all $r$ (where the function $g(r)$ depends only on $\G'$).
            Now, \myitem{belacon} directly follows from \autoref{thm:maindirection}.

        \item \myitem{belacon} $\Longrightarrow$ \myitem{beshrub}. 
            This follows from \autoref{lem:lacontoshrub} and \autoref{def:be}.

        \item \myitem{beshrub} $\Longrightarrow$ \myitem{bebase}. 
            In first-order logic we can compute the (bounded) distance between two vertices,
            check their color and implement the lookup table $S$.
            Thus, we can easily construct a transduction $\tau$ with $\tau(\G') \subseteq \G$.

        \item \myitem{belacon} $\Longrightarrow$ \myitem{beparity}.
            This follows from \autoref{lem:lacontoparity} and \autoref{def:be}.
        \item \myitem{beparity} $\Longrightarrow$ \myitem{bebase}.
            Since the target-degree is bounded, first-order logic can check the parity
            of the intersection of two neighborhoods.\qedhere
    \end{itemize}
\end{proof}

We proceed in a similar way to prove our characterizations of structurally bounded treedepth and treewidth.

\begin{proof}[Proof of \autoref{thm:TDcharacterization}]
    The implications
    \myitem{tdbase} $\Longrightarrow$ \myitem{tdlacon} $\Longrightarrow$ \myitem{tdshrub} $\Longrightarrow$ \myitem{tdbase}
    and then \myitem{tdlacon} $\Longrightarrow$ \myitem{tdparity} $\Longrightarrow$ \myitem{tdbase}
    can be proved analogously to the proof of \autoref{thm:BEcharacterization}.
    We only list the first implication.
    \begin{itemize}
        \item \myitem{tdbase} $\Longrightarrow$ \myitem{tdlacon}.
            As mentioned in the introduction, the weak coloring numbers converge to treedepth, i.e., $\wcol_1(G) \le \dots \le \wcol_\infty(G) = \td(G)$ \cite{coloringcovering}.
            Assume every graph in $\G'$ has treedepth at most $d'$.
            Thus, for a graph $D \in \G$ there exists a graph $G \in \G'$ with $D \in \tau(G)$ and $\wcol_\infty(G) \le d'$.
            \autoref{thm:maindirection} then gives us a lacon-decomposition $(L,\pi)$ of $D$ with
            $\wcol_\infty(L,\pi) \le f(\col_{c}(G)) \cdot \wcol_{\infty}(G) \le f(d') \cdot d'$.\qedhere
    \end{itemize}
\end{proof}

\begin{proof}[Proof of \autoref{thm:TWcharacterization}]
The proof proceeds as the one of \autoref{thm:BEcharacterization} and \ref{thm:TDcharacterization}, except this time
    using $\col_1(G) \le \dots \le \col_\infty(G) = \tw(G)+1$~\cite{coloringcovering}.
\end{proof}

\section{Proof of Localized Feferman--Vaught Composition Theorem}\label{sec:localfv}

We give a self-contained proof of~\autoref{thm:localfv}, using central ideas from the proof of Gaifman's theorem~\cite{grohe2008logic}.
An alternative proof exists in~\cite[Lemma 15]{pilipczuk2018number}.
We repeat the relevant definitions and then prove the result.

\defqtype*

\defseparator*

\thmlocalfv*
\begin{proof}
    Our result follows from proving the following claim via structural induction.
    Let $\phi(\bx, \by_1,\dots,\by_k)$ be a first-order formula with quantifier rank $q$.
    Then one can compute a boolean combination $\Phi(\bx, \by_1, \dots, \by_k)$ of first-order formulas
    of the form $\xi(\bx, \by_i)$ such that 
    for all graphs $G$ and all tuples $\bv_1, \dots, \bv_k$ that are $4^q$-separated by a tuple $\bu$ in $G$ holds
    \[
    G \models \phi(\bu,\bv_1,\dots,\bv_k) \iff G \models \Phi(\bu,\bv_1,\dots,\bv_k).
    \]
    We call such a boolean combination $\Phi$ a $\emph{separated expression}$.
    The claim holds for atomic formulas because $\bu$ $4^0$-separates
    $\bv_1,\dots,\bv_k$, i.e., there are no edges between $\elements(\bv_i)\setminus
    \elements(\bu)$ and $\elements(\bv_j)\setminus \elements(\bu)$ in $G$ for $i \neq j$.
    It also holds for boolean combinations and negations.
    It remains to prove our claim for the case that $\phi(\bx,\by_1,\dots,\by_k) = \exists x \psi$ for some formula~$\psi$.

    For $t \in \N$ and $v \in V(G)$, we define $N^\bu_t(v)$ to be the set of all vertices in $G$
    that are reachable from $v$ via a path of length at most $t$ that contains no vertex from $\bu$.
    For a tuple $\bv$, we define $N^\bu_t(\bv) = \bigcup_{v \in \elements(\bv)} N^\bu_t(v)$.
    We write $z \in N^\bx_t(\by)$ as a short-hand for a first-order formula $\nu_t(\bx,\by,z)$
    such that for all graphs $G$, vertex-tuples $\bu$, $\bv$ and vertices $w$ holds
    $G \models \nu_t(\bu,\bv,w)$ iff $w \in N_t^\bu(\bv)$.
    The formula $\nu_t(\bx,\by,z)$ can be constructed using $t-1$ existential quantifiers.

    Let $r = 4^{q-1}$. 
    We rewrite $\phi$ by distinguishing two cases: Either the existentially quantified variable $x$
    is contained in the neighborhood $N^\bx_r(\by_i)$ for some $i$ or for none.
    This gives~us
    \begin{equation}\label{eq:psiipsistar}
        \phi(\bx,\by_1,\dots,\by_k) \equiv
        \phi^*(\bx,\by_1,\dots,\by_k)
        \lor
        \bigvee_{i=1}^k \phi_i(\bx,\by_1,\dots,\by_k) 
    \end{equation}
    with
    \begin{align*}
        \phi_i(\bx,\by_1,\dots,\by_k) &= \exists x \; x \in N^\bx_r(\by_i) \land \psi(\bx, x, \by_1,\dots,\by_k), \\
        \phi^*(\bx,\by_1,\dots,\by_k) &= \exists x \; x \not\in \bigcup_{i=1}^k N^\bx_r(\by_i) \land \psi(\bx,x,\by_1,\dots,\by_k).
    \end{align*}

    We will proceed by finding separated expressions for $\phi^*$ and for each $\phi_i$.
    Since $4^q = 4 r$, we consider a graph $G$ and tuples $\bv_1, \dots, \bv_k$ that are $4 r$-separated by a tuple $\bu$ in $G$.
    We start with $\phi_i$.
    This formula asks whether there exists an $x \in N^\bx_r(\by_i)$ satisfying $\psi$.
    For every $u \in N^\bu_r(\bv_i)$ holds that 
    $\bu$ $r$-separates the tuples $\bv_i u$ and $\bv_j$ for $i \neq j$,
    since otherwise there would be a short path from $\bv_i$ via $u$ to $\bv_j$ that contains no vertex from $\bu$.
    By the induction hypothesis, there exists a separated expression $\Psi_i(\bx, \by_1, \dots, \by_i x, \dots \by_k)$ such that
    for all $u \in N^\bu_r(\bv_i)$
    \[
    G \models \psi(\bu, u,\bv_1,\dots,\bv_k) \iff G \models \Psi_i(\bu, \bv_1, \dots, \bv_i u, \dots \bv_k).
    \]
    Thus $G \models \phi_i(\bu,\bv_1,\dots,\bv_k) \iff G \models \exists x \; x \in N^\bu_r(\bv_i) \land \Psi_i(\bu, \bv_1, \dots, \bv_ix,\dots \bv_k)$.
    If we assume $\Psi_i$ to be of the form
    \[
    \Psi_i = \bigvee_{l=1}^m \Bigl( \xi_{li}(\bx,\by_i x) \land \bigwedge_{\substack{j=1 \\ j\neq i}}^k \xi_{lj}(\bx,\by_j) \Bigr),
    \]
    then 
    $G \models \phi_i(\bu,\bv_1,\dots,\bv_k) \iff G \models \Phi_i(\bu, \bv_1, \dots, \bv_k)$
    with the separated expression 
    \[
        \Phi_i (\bx, \by_1, \dots, \by_k)= 
        \bigvee_{l=1}^m \Bigl( \exists x \; x \in N^\bx_r(\by_i) \land \xi_{li}(\bx,\by_i x)
        \land \bigwedge_{\substack{j=1 \\ j\neq i}}^k \xi_{lj}(\bx,\by_j) \Bigr).
    \]

    Next, we want to obtain a separated expression for $\phi^*$.
    Let therefore $u \in V(G)$ with $u \not\in N^\bu_r(\bv_i)$ for all $i$.
    Then the tuples $u,\bv_1,\dots,\bv_k$ are $r$-separated by $\bu$.
    By the induction hypothesis, there exists a separated expression $\Psi^*(\bx, x, \by_1, \dots, \by_k)$ such that
    \[
    G \models \psi(\bu, u,\bv_1,\dots,\bv_k) \iff G \models \Psi^*(\bu, u, \bv_1, \dots, \bv_k).
    \]
    We can assume $\Psi^*$ to be of the form
    \[
    \Psi^* = \bigvee_{l=1}^m \Bigl( \xi_{l}(\bx,x) \land \bigwedge_{j=1}^k \xi_{lj}(\bx,\by_j) \Bigr).
    \]
    Then $G \models \phi^*(\bu,\bv_1,\dots,\bv_k) \iff G \models \Phi^*(\bu, \bv_1, \dots, \bv_k)$,
    where
    \[
    \Phi^*(\bx, \by_1, \dots, \by_k)=
    \bigvee_{l=1}^m \Bigl(
    \exists x \; x \not\in \bigcup_{i=1}^k N^\bu_r(\bv_i) \land \xi_l(\bx,x)
    \land
    \bigwedge_{j=1}^k \xi_{lj}(\bx,\by_j) \Bigr).
    \]
    Note that $\Phi^*$ is not yet a separated expression.
    Nevertheless,
    substituting $\Phi_i$ and $\Phi^*$ into equation (\ref{eq:psiipsistar}) yields
    \begin{multline*}
        G \models \phi(\bu,\bv_1,\dots,\bv_k) \iff \\ G \models \bigvee_{i=1}^k \Phi_i(\bu,\bv_1,\dots,\bv_k)
        \lor \bigvee_{l=1}^m \Bigl(
        \exists x \; x \not\in \bigcup_{i=1}^k N^\bu_r(\bv_i) \land \xi_l(\bu,x)
        \bigwedge_{j=1}^k \xi_{lj}(\bu,\bv_j) \Bigr).
    \end{multline*}
    The remaining problematic subformulas are those of the form
    \[
        \gamma(\bx,\by_1,\dots,\by_k) = \exists x \; x \not\in \bigcup_{i=1}^k N^\bx_r(\by_i) \land \xi(\bx,x).
    \]
    To complete the proof, it is sufficient to find a separated expression equivalent to $\gamma$.
    We therefore define the separated expression
    \[
    \Gamma(\bu,\bv_1,\dots,\bv_k) = \bigvee_{i=1}^{k} \exists x \; x \in N^\bu_{3r}(\bv_i) \land x \not\in N^\bu_r(\bv_i) \land \xi(\bu,x)
    \]
    and make a case distinction based on it.

    \begin{itemize}
        \item
        \emph{Case 1:} $G \models \Gamma(\bu,\bv_1,\dots,\bv_k)$.
        Since $\bv_1,\dots,\bv_k$ are $4 r$-separated by $\bu$,
        we have $N^\bu_{3r}(\bv_i) \cap N^\bu_{r}(\bv_j) = \emptyset$ for $i \neq j$.
        This means $G \models \gamma(\bu,\bv_1,\dots,\bv_k)$.

        \item
        \emph{Case 2:} $G \not\models \Gamma(\bu,\bv_1,\dots,\bv_k)$.
        We define the \emph{$\bu$-distance} between two vertices to be the length of the shortest path
        between them that contains no vertex from $\bu$.
        For $V' \subseteq V(G)$, we define an \emph{$(2r,\bu)$-scattered subset} of $V'$ to be a
        set $S \subseteq V'$ such that all vertices in $S$ pairwise have $\bu$-distance greater than $2r$ in $G$.
        The size of the largest $(2r,\bu)$-scattered subset of $V'$ in $G$ is denoted by $s(V')$.
        Let further $R(\bu) = \{ u \mid u \in V(G), G \models \xi(\bu, u) \}$.
        Using this new notation, we observe that $G \models \gamma(\bu,\bv_1,\dots,\bv_k)$ if and only if
        $R(\bu) \setminus \bigcup_{i=1}^k N^\bu_r(\bv_i) \neq \emptyset$.
        Since we assume $G \not\models \Gamma(\bu,\bv_1,\dots,\bv_k)$,
        this means that the sets 
        $R(\bu) \setminus \bigcup_{i=1}^k N^\bu_{r}(\bv_i)$, $R(\bu) \cap N^\bu_r(\bv_1), \dots, R(\bu) \cap N^\bu_r(\bv_k)$
        are $2r$-separated by $\bu$.
        Therefore
        \[
        s\bigl(R(\bu)\bigr) = s\bigl(R(\bu) \setminus \bigcup_{i=1}^k N^\bu_{r}(\bv_i)\bigr) + \sum_{i=1}^k s\bigl(R(\bu) \cap N^\bu_r(\bv_i)\bigr).
        \]
    \end{itemize}
    Thus,
    $G \models \gamma(\bu,\bv_1,\dots,\bv_k)$ iff
    $R(\bu) \setminus \bigcup_{i=1}^k N^\bu_r(\bv_i) \neq \emptyset$ iff
    $s\bigl(R(\bu) \setminus \bigcup_{i=1}^k N^\bu_r(\bv_i)\bigr) \neq 0$ iff
    $\sum_{i=1}^k s\bigl(R(\bu) \cap N^\bu_r(\bv_i)\bigr) < s\bigl(R(\bu)\bigr)$.
    The previous case distinction implies that
    \[
        G \models \gamma(\bu,\bv_1,\dots,\bv_k) \iff
        G \models \Gamma(\bu,\bv_1,\dots,\bv_k) 
        \textnormal{ or }
        \sum_{i=1}^k s\bigl(R(\bu) \cap N^\bu_r(\bv_i)\bigr) < s\bigl(R(\bu)\bigr).
    \]

    It now remains to construct a separated expression $\Delta(\bx,\by_1,\dots,\by_k)$ such that
    $G \models \Delta(\bu,\bv_1,\dots,\bv_k) \iff
    \sum_{i=1}^k s\bigl(R(\bu) \cap N^\bu_r(\bv_i)\bigr) < s\bigl(R(\bu)\bigr)$.
    For arbitrary vertices $a,b,c$ such that $a,b \in N^\bu_r(c)$, we see that $a,b$ have $\bu$-distance at most $2r$.
    Since $s\bigl(R(\bu) \cap N^\bu_r(\bv_i)\bigr)$ is the size of a subset of $N^\bu_r(\bv_i)$ where all vertices
    pairwise have $\bu$-distance greater than $2r$, we have
    $s\bigl(R(\bu) \cap N^\bu_r(\bv_i)\bigr) \le |\bv_i|$.

    For $1 \le i \le k$ and $0 \le h \le |\bv_i|$ we define a first-order formula $\delta_{ih}(\bx,\by_i)$
    which is true iff $s\bigl(R(\bx) \cap N^\bx_r(\by_i)\bigr) \le h$:
    \[
    \delta_{ih}(\bx,\by_i) = 
    \neg \exists s_1 \dots \exists s_{h+1}
            \bigwedge_{j=1}^{h+1}
            \Bigl( 
            \xi(\bx,s_j)
            \land
            s_j \in N^\bx_r(\by_i)
            \land
            \bigwedge_{\substack{l=j+1}}^{h+1}
            s_j \not\in N^\bx_{2r}(s_l)
        \Bigr).
    \]
    We further define for $0 \le h \le \sum_{i=1}^k|\bv_i|$
    a first-order formula $\delta_{h}(\bx)$ which is true iff $s\bigl(R(\bx)\bigr) > h$.
    This formula can be constructed in a similar way as the formulas $\delta_{ih}(\bx,\by_i)$ above.
    We use these formulas to construct the separated expression 
    \[
        \Delta(\bx,\by_1,\dots,\by_k) =
        \bigvee_{h_1 = 0}^{|v_1|}
        \dots
        \bigvee_{h_k = 0}^{|v_k|}
        \delta_{1h_1}(\bx,\by_1)
        \land \dots \land
        \delta_{kh_k}(\bx,\by_k)
        \land \delta_{\sum_{i=1}^k h_i}(\bx)
    \]
    with
    $G \models \Delta(\bu,\bv_1,\dots,\bv_k) \iff \sum_{i=1}^k s\bigl(R(\bu) \cap N^\bu_r(\bv_i)\bigr) < s\bigl(R(\bu)\bigr)$.
    This completes our decomposition of $\phi$ into separated expressions.
\end{proof}

\bibliographystyle{alphaurl}
\bibliography{references}

\newcommand{\etalchar}[1]{$^{#1}$}
\begin{thebibliography}{dMON{\etalchar{+}}19}

\bibitem[BDG{\etalchar{+}}22]{mccw}
\'{E}douard Bonnet, Jan Dreier, Jakub Gajarsk\'{y}, Stephan Kreutzer, Nikolas
  M\"{a}hlmann, Pierre Simon, and Szymon Toru\'{n}czyk.
\newblock Model checking on interpretations of classes of bounded local
  cliquewidth.
\newblock In {\em Proceedings of the 37th Annual ACM/IEEE Symposium on Logic in
  Computer Science ({LICS} 2022)}, New York, NY, USA, 2022. Association for
  Computing Machinery.
\newblock \href {https://doi.org/10.1145/3531130.3533367}
  {\path{doi:10.1145/3531130.3533367}}.

\bibitem[BKTW20]{twinwidth1}
{\'{E}}douard Bonnet, Eun~Jung Kim, St{\'{e}}phan Thomass{\'{e}}, and
  R{\'{e}}mi Watrigant.
\newblock Twin-width {I:} tractable {FO} model checking.
\newblock In {\em 2020 IEEE 61st Annual Symposium on Foundations of Computer
  Science (FOCS)}, pages 601--612, 2020.
\newblock \href {https://doi.org/10.1109/FOCS46700.2020.00062}
  {\path{doi:10.1109/FOCS46700.2020.00062}}.

\bibitem[CHKX06]{chen2006strong}
Jianer Chen, Xiuzhen Huang, Iyad~A Kanj, and Ge~Xia.
\newblock Strong computational lower bounds via parameterized complexity.
\newblock {\em Journal of Computer and System Sciences}, 72(8):1346--1367,
  2006.

\bibitem[Cou90]{Courcelle1990}
Bruno Courcelle.
\newblock The monadic second-order logic of graphs {I}. {R}ecognizable sets of
  finite graphs.
\newblock {\em Inf. Comput.}, 85(1):12--75, 1990.
\newblock \href {https://doi.org/10.1016/0890-5401(90)90043-H}
  {\path{doi:10.1016/0890-5401(90)90043-H}}.

\bibitem[DGK{\etalchar{+}}22]{dreier2022treelike}
Jan Dreier, Jakub Gajarský, Sandra Kiefer, Micha\l{} Pilipczuk, and Szymon
  Toruńczyk.
\newblock Treelike decompositions for transductions of sparse graphs.
\newblock 2022.
\newblock \href {https://doi.org/10.1145/3531130.3533349}
  {\path{doi:10.1145/3531130.3533349}}.

\bibitem[DKR20]{fomc}
Jan Dreier, Philipp Kuinke, and Peter Rossmanith.
\newblock First-order model-checking in random graphs and complex networks.
\newblock In {\em 28th Annual European Symposium on Algorithms (ESA 2020)},
  volume 173 of {\em LIPIcs}. Schloss Dagstuhl - Leibniz-Zentrum f{\"{u}}r
  Informatik, 2020.

\bibitem[DKT10]{dvorak2010deciding}
Zden{\v{e}}k Dvo{\v{r}}{\'a}k, Daniel Kr{\'{a}}l$\!$', and Robin Thomas.
\newblock {Deciding First-Order Properties for Sparse Graphs}.
\newblock In {\em Proceedings of the 51st Conference on Foundations of Computer
  Science}, pages 133--142, 2010.

\bibitem[DKT13]{DvorakKT2013}
Zden{\v{e}}k Dvo{\v{r}}{\'a}k, Daniel Kr{\'{a}}l$\!$', and Robin Thomas.
\newblock Testing first-order properties for subclasses of sparse graphs.
\newblock {\em J. {ACM}}, 60(5):36:1--36:24, 2013.
\newblock \href {https://doi.org/10.1145/2499483} {\path{doi:10.1145/2499483}}.

\bibitem[dMON{\etalchar{+}}19]{de2019shrub}
Patrice~Ossona de~Mendez, Jan Obdr{\v{z}}{\'a}lek, Jaroslav
  Ne{\v{s}}et{\v{r}}il, Petr Hlin{\v{e}}n{\`y}, and Robert Ganian.
\newblock Shrub-depth: Capturing height of dense graphs.
\newblock {\em Logical Methods in Computer Science}, 15, 2019.

\bibitem[DMS23]{mcsnd}
Jan Dreier, Nikolas Mählmann, and Sebastian Siebertz.
\newblock First-order model checking on structurally sparse graph classes,
  2023.
\newblock \href {https://doi.org/10.48550/ARXIV.2302.03527}
  {\path{doi:10.48550/ARXIV.2302.03527}}.

\bibitem[FG01a]{FlumG2001}
J{\"o}rg Flum and Martin Grohe.
\newblock {Fixed-Parameter Tractability, Definability, and Model-Checking}.
\newblock {\em SIAM Journal on Computing}, 31(1):113--145, 2001.

\bibitem[FG01b]{FrickG2001}
Markus Frick and Martin Grohe.
\newblock Deciding first-order properties of locally tree-decomposable
  structures.
\newblock {\em J. {ACM}}, 48(6):1184--1206, 2001.
\newblock \href {https://doi.org/10.1145/504794.504798}
  {\path{doi:10.1145/504794.504798}}.

\bibitem[Gai82]{gaifman1982local}
Haim Gaifman.
\newblock On local and non-local properties.
\newblock In {\em Studies in Logic and the Foundations of Mathematics}, volume
  107, pages 105--135. Elsevier, 1982.

\bibitem[GHN{\etalchar{+}}12]{originalshrubdepth}
Robert Ganian, Petr Hlin{\v{e}}n{\'y}, Jaroslav Ne{\v{s}}et{\v{r}}il, Jan
  Obdr{\v{z}}{\'a}lek, Patrice Ossona~de Mendez, and Reshma Ramadurai.
\newblock When trees grow low: Shrubs and fast mso1.
\newblock In {\em Mathematical Foundations of Computer Science 2012}, pages
  419--430, Berlin, Heidelberg, 2012. Springer Berlin Heidelberg.

\bibitem[GHO{\etalchar{+}}16]{gajarsky2016new}
Jakub Gajarsk{\'y}, P~Hlin{\v{e}}n{\'y}, J~Obdr{\v{z}}{\'a}lek, Daniel
  Lokshtanov, and MS~Ramanujan.
\newblock {A New Perspective on FO Model Checking of Dense Graph Classes}.
\newblock In {\em Proceedings of the 31st Symposium on Logic in Computer
  Science}, pages 176--184, 2016.

\bibitem[GJdM{\etalchar{+}}21]{no-qe}
Mario Grobler, Yiting Jiang, Patrice~Ossona de~Mendez, Sebastian Siebertz, and
  Alexandre Vigny.
\newblock Discrepancy and sparsity, 2021.
\newblock \href {https://doi.org/10.48550/ARXIV.2105.03693}
  {\path{doi:10.48550/ARXIV.2105.03693}}.

\bibitem[GKN{\etalchar{+}}20]{gajarsky2018first}
Jakub Gajarsk\'{y}, Stephan Kreutzer, Jaroslav Ne\v{s}et\v{r}il, Patrice
  Ossona~De Mendez, Micha\l{} Pilipczuk, Sebastian Siebertz, and Szymon
  Toru\'{n}czyk.
\newblock First-order interpretations of bounded expansion classes.
\newblock {\em ACM Trans. Comput. Logic}, 21(4), July 2020.
\newblock \href {https://doi.org/10.1145/3382093} {\path{doi:10.1145/3382093}}.

\bibitem[GKR{\etalchar{+}}18]{coloringcovering}
Martin Grohe, Stephan Kreutzer, Roman Rabinovich, Sebastian Siebertz, and
  Konstantinos~S. Stavropoulos.
\newblock Coloring and covering nowhere dense graphs.
\newblock {\em {SIAM} J. Discret. Math.}, 32(4):2467--2481, 2018.
\newblock \href {https://doi.org/10.1137/18M1168753}
  {\path{doi:10.1137/18M1168753}}.

\bibitem[GKS17]{GroheKS2017}
Martin Grohe, Stephan Kreutzer, and Sebastian Siebertz.
\newblock Deciding first-order properties of nowhere dense graphs.
\newblock {\em J. {ACM}}, 64(3):17:1--17:32, 2017.
\newblock \href {https://doi.org/10.1145/3051095} {\path{doi:10.1145/3051095}}.

\bibitem[Gro08]{grohe2008logic}
Martin Grohe.
\newblock Logic, graphs, and algorithms.
\newblock {\em Logic and Automata}, 2:357--422, 2008.

\bibitem[Han65]{Hanf1965}
William Hanf.
\newblock Model-theoretic methods in the study of elementary logic.
\newblock In {\em Journal of Symbolic Logic}, pages 132--145. Amsterdam:
  North-Holland Pub. Co., 1965.

\bibitem[Kar67]{FV59}
Carol Karp.
\newblock The first order properties of products of algebraic systems.
  fundamenta mathematicae.
\newblock {\em Journal of Symbolic Logic}, 32(2):276–276, 1967.
\newblock \href {https://doi.org/10.2307/2271704} {\path{doi:10.2307/2271704}}.

\bibitem[KS17]{KuskeS2017}
Dietrich Kuske and Nicole Schweikardt.
\newblock First-order logic with counting.
\newblock In {\em 32nd Annual {ACM/IEEE} Symposium on Logic in Computer
  Science, {LICS} 2017, Reykjavik, Iceland, June 20-23, 2017}, pages 1--12.
  {IEEE} Computer Society, 2017.
\newblock \href {https://doi.org/10.1109/LICS.2017.8005133}
  {\path{doi:10.1109/LICS.2017.8005133}}.

\bibitem[KY03]{coloringdefinition}
Henry~A. Kierstead and Daqing Yang.
\newblock Orderings on graphs and game coloring number.
\newblock {\em Order}, 20(3):255--264, 2003.
\newblock \href {https://doi.org/10.1023/B:ORDE.0000026489.93166.cb}
  {\path{doi:10.1023/B:ORDE.0000026489.93166.cb}}.

\bibitem[Mak04]{makowsky2004algorithmic}
Johann~A Makowsky.
\newblock Algorithmic uses of the {F}eferman--{V}aught theorem.
\newblock {\em Annals of Pure and Applied Logic}, 126(1-3):159--213, 2004.

\bibitem[NdM12]{NesetrilM2012}
Jaroslav Ne\v{s}et\v{r}il and Patrice~Ossona de~Mendez.
\newblock {\em Sparsity - Graphs, Structures, and Algorithms}, volume~28 of
  {\em Algorithms and combinatorics}.
\newblock Springer, 2012.
\newblock \href {https://doi.org/10.1007/978-3-642-27875-4}
  {\path{doi:10.1007/978-3-642-27875-4}}.

\bibitem[NDM15]{nevsetvril2015low}
Jaroslav Ne{\v{s}}et{\v{r}}il and Patrice~Ossona De~Mendez.
\newblock On low tree-depth decompositions.
\newblock {\em Graphs and combinatorics}, 31(6):1941--1963, 2015.

\bibitem[NdMS20]{nesetril2020towards}
Jaroslav Ne\v{s}et\v{r}il, Patrice~Ossona de~Mendez, and Sebastian Siebertz.
\newblock Towards an arboretum of monadically stable classes of graphs.
\newblock {\em arXiv preprint arXiv:2010.02607}, 2020.

\bibitem[PST18]{pilipczuk2018number}
Micha{\l} Pilipczuk, Sebastian Siebertz, and Szymon Toru{\'n}czyk.
\newblock On the number of types in sparse graphs.
\newblock In {\em Proceedings of the 33rd Annual ACM/IEEE Symposium on Logic in
  Computer Science}, pages 799--808, 2018.

\bibitem[See96]{Seese1996}
Detlef Seese.
\newblock Linear time computable problems and first-order descriptions.
\newblock {\em Math. Struct. Comput. Sci.}, 6(6):505--526, 1996.

\bibitem[vK21]{van2019uniform}
Jan {van den Heuvel} and H.A. Kierstead.
\newblock Uniform orderings for generalized coloring numbers.
\newblock {\em European Journal of Combinatorics}, 91:103214, 2021.
\newblock Colorings and structural graph theory in context (a tribute to Xuding
  Zhu).
\newblock \href {https://doi.org/10.1016/j.ejc.2020.103214}
  {\path{doi:10.1016/j.ejc.2020.103214}}.

\bibitem[Zhu09]{zhu2009colouring}
Xuding Zhu.
\newblock Colouring graphs with bounded generalized colouring number.
\newblock {\em Discrete Mathematics}, 309(18):5562--5568, 2009.

\end{thebibliography}
\end{document}